\documentclass[usletter,12pt]{article}
\pdfoutput=1 
\usepackage{jheppub} 
\usepackage[T1]{fontenc} 
\usepackage{amsfonts,amssymb,amsmath}
\usepackage{color}

\newcommand{\be}{\begin{equation}}
\newcommand{\ee}{\end{equation}}
\newcommand{\bea}{\begin{eqnarray}}
\newcommand{\eea}{\end{eqnarray}}
\newcommand{\beal}{\begin{aligned}}
\newcommand{\eeal}{\end{aligned}}

\title{Vacuum metastability with black holes}

\author[a]{Philipp Burda\footnote{On leave of absence from ITEP, Moscow.}}
\author[a,b]{Ruth Gregory}
\author[c]{Ian G.\ Moss}

\affiliation[a]{Centre for Particle Theory, Durham University,
South Road, Durham, DH1 3LE, UK}
\affiliation[b]{Perimeter Institute, 31 Caroline Street North, Waterloo, 
ON, N2L 2Y5, Canada}
\affiliation[c]{School of Mathematics and Statistics, Newcastle University, 
Newcastle Upon Tyne, NE1 7RU, U.K.}

\emailAdd{philipp.burda@durham.ac.uk}
\emailAdd{r.a.w.gregory@durham.ac.uk}
\emailAdd{ian.moss@newcastle.ac.uk}

\abstract{
We consider the possibility that small black holes can act as 
nucleation seeds for the decay of a metastable vacuum, 
focussing particularly on the Higgs potential. 
Using a thin-wall bubble approximation for the nucleation 
process, which is possible when generic quantum gravity 
corrections are added to the Higgs potential,
we show that primordial black holes can stimulate vacuum 
decay. We demonstrate that for suitable parameter
ranges, the vacuum decay process dominates over 
the Hawking evaporation process. Finally, we comment on the
application of these results to vacuum decay seeded by black holes
produced in particle collisions.}

\preprint{DCPT-15/11}

\begin{document}

\maketitle

\section{Introduction}

We live in a world in which the fundamental properties of matter 
are manifestly unchanging on the timescale of our everyday lives.
Nevertheless, the recent discovery of the Higgs boson 
\cite{ATLAS:2012ae,Chatrchyan:2012tx} raises the possibility
that, even within the standard model of particle physics, 
the present vacuum state of the universe may not be stable,
but only metastable, with another lower energy state at high
expectation values of the Higgs field \cite{Degrassi:2012ry,
Gorsky:2014una,Bezrukov:2014ina,Ellis:2015dha,Blum:2015rpa}.
In general, this would not conflict with observation because the lifetime of the 
present vacuum would be far longer than the age of the universe.
Indeed, the possibility that we live in a metastable state was mooted long before 
the discovery of the Higgs
\cite{Krive:1976sg,Cabibbo:1979ay,1982Natur.298,
Lindner:1988ww,Sher:1988mj,Isidori:2001bm,Espinosa:2007qp,
Isidori:2007vm,EliasMiro:2011aa}.

Investigations of vacuum decay in the context of quantum field 
theory are usually based on the bubble nucleation arguments of 
Coleman et al.\ \cite{coleman1977,callan1977,CDL}, (see also
\cite{Kobzarev:1974cp}) which relate vacuum decay to the 
random nucleation of critical bubbles of 
a new vacuum or phase.
However, in many familiar examples of phase transitions 
beyond the realm of particle physics, the transition is dominated 
by bubbles which nucleate around fixed sites, usually impurities in 
the medium or imperfections in a containment vessel. It is therefore
important to investigate whether the metastable Higgs vacuum 
might be ruled out if the seeded nucleation rates
for vacuum decay are comparatively large.

In recent work \cite{GMW}, following earlier work by Hiscock 
and Berezin \cite{PhysRevD.35.1161,Berezin:1987ea},
we looked at the effect of gravitational inhomogeneities acting
as seeds of cosmological phase transitions in de Sitter space.
We found that the decay rates were considerably enhanced 
by the presence of black holes.
Following our work, Sasaki and Yeom \cite{Sasaki:2014spa} 
have investigated the unitarity implications of bubble
nucleation in Schwarzschild-Anti de Sitter spacetimes
(see also \cite{Shkerin:2015exa} for a discussion of
vacuum stability in the early universe).
In this paper we extend our previous results \cite{GMW}, 
to cover all possible gravitational nucleation processes, 
focussing in particular on the nucleation of bubbles of 
Anti de Sitter (AdS) spacetime within a vacuum first reported
in \cite{Burda:2015isa}.

We follow the approach of Coleman and de Luccia \cite{CDL}, 
and assume that the nucleation probability for a bubble of the 
new phase is given schematically by
\begin{equation}
\Gamma=Ae^{-B},
\label{GammaD}
\end{equation}
where $B$ is the action of an imaginary-time solution to the 
Einstein-Higgs field equations, or instanton,  which approaches 
the false vacuum at large distances. However, unlike Coleman 
and de Luccia, we consider a spherically symmetric bubble on 
a black hole background. The nucleation process typically requires 
an instanton that has a conical singularity at the black hole horizon. 
Analogous instantons were considered before in 
\cite{Mellor:1989gi, Mellor:1989wc} and fall within the generalised 
type introduced by Hawking and Turok \cite{Hawking:1998bn, Turok:1998he}. 
As in our previous paper, we show that the nucleation probability is 
well-defined. An alternative interpretation of (\ref{GammaD}) and 
the instanton has been given in \cite{Brown:2007sd}.

The vacuum decay process is based on a static black hole, 
in which a bubble nucleates outside the black hole and 
either completely replaces the black hole with a bubble
of true vacuum expanding outwards, or nucleates a 
static bubble leaving a remnant black hole surrounded 
by true vacuum. This latter solution is not stable, and
small fluctuations will lead it to either expand as with the
first situation completing the phase transition, or to collapse 
back inwards leaving the initial state unchanged.
Of course, this description does not explicitly account for
any time dependence of the black hole due to Hawking 
evaporation, however, we can apply the same argument 
as that employed for black hole particle production, namely,
we consider only vacuum decay precesses which have 
timescales short compared to the evaporation rate.
In other words, we have some confidence in our results 
when the vacuum decay rate exceeds the mass decay rate 
of the black hole. (The effects of Hawking radiation
on tunnelling rates have been investigated in 
\cite{PhysRevD.32.1333,Cheung:2013sxa}). 

We will show that the vacuum decay seeded by
black holes greatly exceeds the Hawking evaporation
rate for particle physics scale bubbles. This clearly has
relevance for the Higgs potential, which we consider
explicitly in \textsection \ref{higgssec}.
A primordial black hole losing mass by the Hawking process
would decay down to a mass around 10-100 times the Planck 
mass and then seed a vacuum transition. The fact that this 
has not happened therefore means that either the Higgs 
parameters are not the the relevant range (a small region
of parameter space for this purely gravitational argument) 
or there are no primordial black
holes in the observable universe.

Since our main application is to the Higgs vacuum, we
will first summarize some of the features of the Higgs potential
relevant to the calculation. As with the phenomenological
explorations of the Higgs potential, we write the
potential in terms of an overall magnitude of
the Higgs, $\phi$, and approximate the potential
with an effective coupling $\lambda_{\rm eff}$,
\be
V(\phi)=\frac14\lambda_{\rm eff}(\phi)\phi^4.
\ee
The exact form of $\lambda_{\rm eff}$ is determined by a
renormalisation group computation with the parameters and
masses measured at low-energy. Two-loop calculations 
of the running coupling
\cite{Ford:1992mv,Chetyrkin:2012rz,Bezrukov:2012sa,Degrassi:2012ry}, 
can be approximated by an expression of the form 
\be
\lambda_{\rm eff}\approx\lambda_*+b\left(\ln{\frac{\phi}{\phi_*}}\right)^2,
\ee
where $-0.01\lesssim\lambda_*\lesssim0$, 
$0.1M_p\lesssim \phi_*\lesssim M_p$ and $b\sim 10^{-4}$. 
The uncertainty on these parameter ranges is due mostly
to experimental uncertainties in the Higgs mass and the top quark 
mass, however the possibility of negative $\lambda_{\rm eff}$ 
approaching the Planck scale is quite real. The present-day 
broken symmetry vacuum may therefore be a metastable state, but 
quantum tunnelling in the Higgs potential determined by the usual
Coleman de Luccia expressions is very slow, and the lifetime of 
the false vacuum far exceeds the lifetime of the universe. 
\begin{figure}[htb]
\centering
\includegraphics[width=0.7\textwidth]{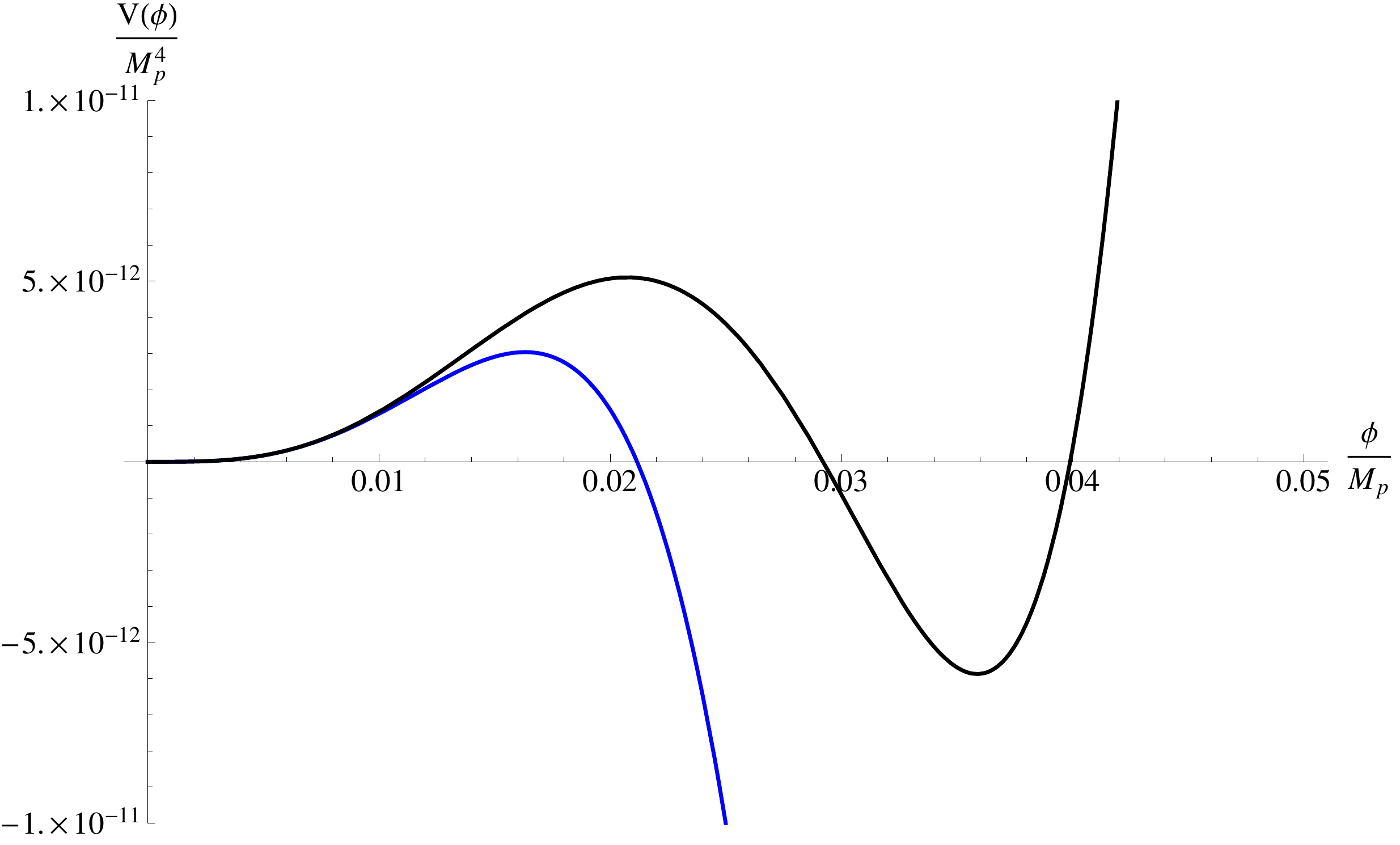}
\caption{
The Higgs potential at large values of one of the Higgs 
field components $\phi$. The parameter values for
the blue line are $\lambda_*=-0.001$, $\phi_*=0.5M_p$. 
The black line shows the effect of adding a $\phi^6$
term with coefficient $\lambda_6=0.34$. }
\label{bsmplot}
\end{figure}

The observation of negative $\lambda_{\rm eff}$ of course 
assumes no corrections from new physics between the TeV 
scale and the Planck scale. 
We might expect quantum gravity, or other effects will have 
to be taken into account.
On dimensional grounds, we can write modifications to the potential 
of the following form 
\cite{Bergerhoff:1999jj, Greenwood:2008qp,Branchina:2013jra,
Branchina:2014rva,Eichhorn2015,Loebbert:2015eea},
\be
V(\phi)=\frac14\lambda_{\rm eff}(\phi)\phi^4
+\frac14(\delta\lambda)_{\rm bsm}\phi^4
+\frac16\lambda_6{\phi^6\over M_p^2}
+\frac18\lambda_8{\phi^8\over M_p^4}+\dots\label{hehiggs}
\ee
where $(\delta\lambda)_{\rm bsm}$ includes corrections from 
BSM physics, and the polynomial terms represent unknown
physics from the Planck scale. If these coefficients are similar in
magnitude, then the small size of $\lambda_{\rm eff}$ at the 
Planck scale has the consequence that there is an intermediate
range of $\phi$ where the potential is determined predominantly 
by $\lambda_{\rm eff}$ and $\lambda_6$.  

Quantum tunnelling in a corrected potential has been explored 
by Branchina et al.\ \cite{Branchina:2013jra,Branchina:2014rva}
(see also \cite{Lalak:2014qua}). 
They considered potentials with $\lambda_*\sim-0.1$, where
the potential barrier occurs at $\phi\ll M_p$, and they further 
enhanced the tunnelling rate by taking $\lambda_6=-2$. They 
claimed a greatly enhanced tunnelling rate, with a lifetime
much shorter than the age of the universe, however, their
discussion did not include gravitational interactions.

The aim of the present paper is to analyse models which appear
stable on cosmological timescales when using the CDL results,
but may become unstable due to enhancement of the tunnelling 
rate by a nucleation seed, which we will take to be a microscopic 
black hole. For this semi-analytic investigation, we consider the 
nucleation with thin-wall bubbles of the true vacuum in an analogous
way to Coleman and de Luccia. In terms of the Higgs, this
thin-wall bubble nucleation requires the potential to be relatively 
shallow at the true vacuum, and this requires a large positive $\phi^6$ 
term. To go beyond this approximation, which allows us to use pure
gravitational arguments, will require a detailed numerical study that
we will present in future work.  

The outline of the paper is as follows. We first review then extend the
thin wall instanton method in \textsection \ref{thinwall}, directly 
calculating the instanton action in the thin wall limit as a function of
wall trajectory and black holes masses. In \textsection 
\ref{Tunnellingsection} we describe the solutions for the instantons 
and discuss the preferred decay process for a general seed mass 
black hole (including charge). In \textsection \ref{higgssec} we apply 
the results to the case of the Higgs potential, and present a full 
comparative calculation with the decay of the black hole due to Hawking
radiation. Finally, in \textsection \ref{discuss}, we discuss
possible extensions to higher dimensions and collider
black holes. Note we use units in
which $\hbar=c=1$, and use the reduced Planck mass 
$M_p^2=1/(8\pi G)$.

\section{Thin-wall bubbles}
\label{thinwall}

In this section we describe how to construct a thin wall instanton, along the
lines of Coleman et al.\ \cite{coleman1977,callan1977,CDL}, 
but with the difference that we suppose that an inhomogeneity is present. 
Complementary to our earlier work \cite{GMW}, we apply Israel's 
thin wall techniques \cite{Israel} to the bubble wall, and describe 
the inhomogeneity by a black hole. (In appendix \ref{altaction} we
calculate the instanton action for more general inhomogeneous
configurations, with the proviso that they be static.)

\subsection{Constructing the instanton}

The physical process of vacuum decay with an inhomogeneity can
be represented gravitationally by a Euclidean solution with two 
`Schwarzschild' bulks which have different cosmological constants 
separated by a thin wall with constant tension (for a general 
proof of this result in the context of braneworlds, see 
\cite{BCG,Gregory:2001xu}).
On each side of the wall the geometry has the form
\begin{equation}
ds^2=f(r)d\tau_\pm^2+{dr^2\over f(r)}+r^2d\Omega_{I\!I}^2,\qquad 
f(r)\equiv 1-\frac{2GM_\pm}{r}-\frac{\Lambda_\pm r^2}{3},
\end{equation}
where $\tau_\pm$ are the different time coordinates on each
side of the wall, and the wall, or boundary of each bulk, is 
parametrised by some trajectory $r=R(\lambda)$ (the angular
$\theta$ and $\phi$ coordinates are the same on each side).
The Israel junction conditions \cite{Israel} relate the solution 
inside the bubble with mass $M_-$ and cosmological constant 
$\Lambda_-$, to the solution outside the bubble with mass $M_+$ 
and cosmological constant $\Lambda_+$. Since the bubble exterior 
is in the false vacuum, we have $\Lambda_+>\Lambda_-$. 
($\Lambda_+<\Lambda_-$ was discussed by Aguirre and Johnson 
\cite{Aguirre:2005xs,Aguirre:2005nt}, and the case $M_-=0$
has been discussed by Sasaki and Yeom \cite{Sasaki:2014spa}).  
In general, the bubble will follow a time-dependent trajectory 
representing a reflection, or bounce.

Following the Israel approach \cite{Israel}, we choose to
parametrize the wall trajectory by the proper time of a 
comoving observer, i.e.\ $\lambda$ is chosen so that
\be
f {\dot \tau}_\pm^2 + \frac{{\dot R}^2}{f} = 1
\ee
and take normal forms that point towards increasing $r$:
\begin{equation}
n_\pm={\dot \tau}_\pm\,dr_\pm-\dot r\,d\tau_\pm,\label{normal}
\end{equation}
where dots denote derivatives with respect to $\lambda$.
We also take ${\dot\tau}_\pm\geq0$ for orientability (see also
\cite{Sasaki:2014spa}).
In these conventions, the Israel junction conditions are
\begin{equation}
f_+\dot t_+-f_-\dot t_-=-4\pi G\sigma R.\label{jnc}
\end{equation}
The combination of surface tension and Newton's constant 
recurs so frequently that for clarity we define
\begin{equation}
{\bar\sigma}=2\pi G\sigma.
\end{equation}

To find solutions to the equations of motion, first note that
the junction condition \eqref{jnc} implies
\begin{equation}
f_\pm\dot \tau_\pm=\left(f_\pm-\dot R^2\right)^{1/2}=
{f_--f_+\over 4\bar\sigma R}\mp\bar\sigma R\,.
\label{junctions}
\end{equation}
It is convenient to rewrite this as an equation for $\dot R$
using the explicit forms for $f_\pm$
\begin{equation}
{\dot R}^2 = 1-\left ( {\bar\sigma}^2 + \frac{\bar\Lambda}{3} 
+ \frac{(\Delta\Lambda)^2}{144{\bar\sigma}^2} \right) R^2 
-\frac{2G}{R} \left ( {\bar M} + \frac{\Delta M \Delta\Lambda}
{24{\bar\sigma}^2}\right) - \frac{(G\Delta M)^2}{4 R^4 {\bar\sigma}^2},
\label{rdots2}
\end{equation}
where $\Delta M=M_+-M_-$ and $\bar M=(M_++M_-)/2$ with similar 
expressions for $\Lambda$.

Although this seems to be a more complex system than that 
considered in \cite{GMW}, in fact it is possible to rescale the
variables so that the analysis is very nearly identical to that
in \cite{GMW}.
To begin with, define
\begin{equation}
\ell^2 = \frac{3}{\Delta\Lambda},\qquad
\gamma = \frac{4{\bar\sigma}\ell^2}{1+4{\bar\sigma}^2\ell^2},\qquad
\alpha^2 = 1 + \frac{\Lambda_- \gamma^2}{3},
\end{equation}
and rescale the coordinates to ${\tilde R}={\alpha R}/{\gamma}$, 
${\tilde \tau} ={\alpha \tau}/{\gamma}$,
${\tilde \lambda}={\alpha \lambda}/{\gamma}$.
Then writing
\begin{equation}
k_1 = {2\alpha GM_-\over\gamma} +{(1-\alpha)\alpha G\Delta M\over \bar\sigma
\gamma^2},
\quad
k_2 = {\alpha^2 G\Delta M\over 2\bar\sigma\gamma^2}.
\end{equation}
gives a Friedman-like equation for ${\tilde R}({\tilde\lambda})$:
\be
\left ( \frac{d{\tilde R}}{d{\tilde \lambda}} \right ) ^2=1
-\left ( {\tilde R}+ \frac{k_2}{{\tilde R}^2} \right)^2 
-\frac{k_1}{\tilde R}=-U(\tilde R)
\label{rdot}
\ee
together with equations for ${\tilde\tau}_\pm$ (given
in appendix \ref{kappalimits}).
These equations with $\alpha=1$ are precisely the system 
explored in \cite{GMW}. The allowed parameter ranges for 
$k_1$ and $k_2$ are obtained similarly, and
discussed in appendix \ref{kappalimits}.

\subsection{Computing the action}
\label{actioncomputation}

To compute the action of the bounce, we need to compute the 
Euclidean action of the thin wall instanton:
\be
\beal
I_E = &-\frac{1}{16\pi G} \int_{{\cal M}_+} \sqrt{g} 
({\cal R}_+-2\Lambda_+)
-\frac{1}{16\pi G} \int_{{\cal M}_-} \sqrt{g} 
({\cal R}_--2\Lambda_-)\\
&+ \frac{1}{8\pi G} \int_{\partial{\cal M}_+} \sqrt{h} K_+ 
- \frac{1}{8\pi G} \int_{\partial{\cal M}_-} \sqrt{h} K_- 
+ \int_{\cal W} \sigma \sqrt{h}
\eeal
\ee
and subtract the action of the background. In this expression,
$\partial{\cal M}_\pm$ refers to the boundary induced by the
wall -- there may also be additional boundary or bulk terms
required for renormalisation of the action (see below).
Note that we have reversed the sign of the $\partial{\cal M}_-$
normal in the Gibbons-Hawking boundary term so that it agrees
with the {\it outward} pointing normal of the Israel prescription.
On each side of the wall in the bulk we have 
${\cal R}_\pm=4\Lambda_\pm$,
and the Israel equations give $K_+-K_- = -12\pi G \sigma$.

There are three parts to the computation of the action, ${\cal M}_-$,
${\cal M}_+$, and ${\cal W}$.

\medskip
\noindent $\bullet\; {\cal M}_-$

Integrating the bulk term for the ``$-$'' side of the wall has two 
contributions, one from the cosmological constant in the bulk
volume, and a contribution from any conical deficit at the
black hole event horizon, should one exist. A description of how
to deal with conical deficits was given in an appendix of \cite{GMW},
essentially the deficit gives a contribution proportional to the horizon
area times the deficit angle. Supposing that the periodicity of the
Euclidean time coordinate, $\beta$, set by the wall solution, may not
be the same as the natural horizon periodicity, 
$\beta_- = 4\pi r_h / (1-\Lambda_-r_h^2)$, this gives a 
contribution to the action from ${\cal M}_-$ of:
\be
\beal
I_{{\cal M}_-} &= - \frac{\beta_--\beta}{\beta_-} \frac{{\cal A}_-}{4G} - 
\frac{1}{4 G} \int \frac{2\Lambda_-}{3} (R^3-r_h^3)d\tau_-\\
&= - \frac{{\cal A}_-}{4G} + \frac{\beta}{4G}\left [
\frac{{\cal A}_-}{\beta_-} + \frac{2\Lambda_- r_h^3}{3} - 2GM_-
\right ] + \frac{1}{4G} \int d\lambda R^2 f_-' {\dot\tau}_-
\eeal
\ee
where ${\cal A}_-$ is the area of the black hole horizon in ${\cal M}_-$.
Inserting the value of $\beta_-$, and taking into account the 
value of $r_h$, the term in square brackets is identically zero,
and this contribution to the action does not explicitly depend on
the periodicity or indeed any conical deficit angle.

\medskip
\noindent $\bullet \; {\cal M}_+$

The computation of the action of ${\cal M}_+$ is a little
more involved, as different regularisation prescriptions are
needed for the different asymptotics of (A)dS or flat spacetime.

For Schwarzschild de Sitter, the radial coordinate in the static 
patch has a finite range, and terminates at the cosmological 
horizon $r_c$, which has a natural periodicity $\beta_c 
= -4\pi r_c^2/(2GM_+-2\Lambda_+ r_c^3/3)$.
\be
\beal
I_{{\cal M}_+} 
&= -\frac{(\beta_c-\beta){\cal A}_c}{4G\beta_c}-
\frac{1}{4 G} \int \frac{2\Lambda_+}{3} (r_c^3-R^3)d\tau_+\\
&= -\frac{{\cal A}_c}{4G} + \frac{\beta}{4G}\left [
\frac{{\cal A}_c}{\beta_c} - \frac{2\Lambda_+ r_c^3}{3} + 2GM_+
\right ] - \frac{1}{4G} \int d\lambda R^2 f_+' {\dot\tau}_+
\eeal
\ee
where ${\cal A}_c$ is the area of the cosmological event horizon.
Once again, substituting the values of $\beta_c$ and $r_c$
demonstrates that the bracketed term vanishes. For future
reference, we note the value of the background SDS action
at arbitrary periodicity derived in \cite{GMW}:
\be
I_{ESDS} = -\frac{{\cal A}_c}{4G} -\frac{{\cal A}_+}{4G}
\ee
where ${\cal A}_+$ is the horizon area of the black hole
of mass $M_+$. Note that this expression is $\beta-$independent
as discussed in \cite{GMW}.

For Schwarzschild (and Schwarzschild-AdS) the range of the radial 
coordinate is now infinite, and we must perform a renormalization 
procedure. For Schwarzschild, there is no contribution from the
bulk integral, and instead we consider an artificial boundary at
large $r_0$, with a subtracted Gibbons-Hawking term so that
flat space has zero action \cite{gibbonshawking}.
\be
I_{{\cal M}_+} = \frac{1}{8\pi G} \int_{r=r_0} \sqrt{h} (K-K_0) 
=\frac{\beta M_+}{2} = \beta M_+ - \frac{1}{4G} \int d\lambda f_+'R^2
{\dot\tau}_+
\label{schasympt}
\ee
Again for future reference, computing the background Euclidean
Schwarzschild action at arbitrary periodicity (with the same
background subtraction prescription) yields
\be
I_{ESCH} =  -\frac{{\cal A}_+}{4G}  + \beta M_+
\label{schback}
\ee
inputting the value of $\beta_{SCH}=8\pi GM_+$.

For AdS on the other hand, we must subtract off the divergent
volume contribution \cite{Witten:1998qj} by again introducing
a fiducial boundary at $r_0$, and subtracting a pure AdS
integral, which must have an adjusted time-periodicity so that 
the boundary manifolds at $r_0$ agree:
\be
\beta_0 = \beta \frac{f_+^{1/2}}{(1-\Lambda_+ r_0^2/3)^{1/2} }
\simeq \left (1 + \frac{3GM_+}{r_0^3\Lambda} \right )\beta.
\label{betaadsren}
\ee
Thus
\be
\beal
I_{{\cal M}_+} &=
-\frac{1}{4G} \int d\tau \frac{2\Lambda_+}{3} (r_0^3-R^3)
+\frac{1}{4G} \int d\tau_0 \frac{2\Lambda_+}{3} r_0^3\\
&=\beta M_+ - \frac{1}{4G} \int d\lambda f_+'R^2
{\dot\tau}_+
\eeal
\ee
i.e.\ an identical result to the Schwarzschild case \eqref{schasympt}.
Computing the background Euclidean Schwarzschild-AdS action at
arbitrary periodicity we get
\be
I_{ESADS} = -\frac{{\cal A}_+}{4G} +\beta M_+
\label{adsschback}
\ee
again, the same expression as for Schwarzschild, \eqref{schback}.

\medskip
\noindent $\bullet \; {\cal W}$

Finally, the contribution to the action from the wall has a particularly
simple form as the Gibbons-Hawking boundary terms from
the wall come in the combination of the Israel junction conditions. 
We therefore obtain
\be
I_{\cal W} = 
\pm \frac{1}{8\pi G} \int_{\partial{\cal M}_\pm} \sqrt{h} K 
+ \int_{\cal W} \sigma \sqrt{h} = - \int_{\cal W} \frac\sigma2 \sqrt{h}
=\frac{1}{2G} \int d\lambda\,R\left (f_+ {\dot\tau}_+ - f_- {\dot\tau}_-\right)
\ee
having used $f_+ {\dot\tau}_+ - f_- {\dot\tau}_- = -2{\bar\sigma}R$.

Putting all of these results together, we find that the action of the
instanton solution is
\be
I_E = - \frac{{\cal A}_-}{4G} + \frac{1}{2G} \int d\lambda
\left [  (R -3GM_+){\dot\tau}_+ -  (R -3GM_-){\dot\tau}_-\right]
+ \begin{cases}
\beta M_+ & \Lambda_+\leq0 \\
- \displaystyle{\frac{{\cal A}_c}{4G}} & \Lambda_+>0
\end{cases}
\ee
Thus the bounce action, given by subtracting the background 
Schwarzschild/S(A)dS action is:
\begin{equation}
B= {{\cal A}_+\over 4G}-{{\cal A}_-\over 4G} +{1\over 4G}
\oint d\lambda \left\{\left ( 2Rf_+ - R^2 f_+'\right)\dot{\tau}_+
- \left (2Rf_- -R^2 f_-'\right) \dot{\tau}_-\right\}
\label{bounceaction}
\end{equation}
This expression is the central result of this section, and is
independent of any choice of periodicity of Euclidean time,
and independent of the choices of cosmological constant
on each side of the wall. It is in fact even valid when
the black hole is charged, as we will consider in the
next section.

\section{Instanton solutions}
\label{Tunnellingsection}

In the previous section we derived the equations of motion for 
a bubble wall separating a region of true vacuum from the false 
vacuum, and derived the ``master expression'' \eqref{bounceaction}
for the instanton action. In this section we discuss general 
properties of these solutions, and demonstrate how the action
varies as we change the seed black hole mass and the wall
tension. Rather than presenting absolute values of the bounce
action, it proves useful instead to present a comparator to the
`Coleman de Luccia' action, by which we mean the bounce 
solution in the absence of any black holes (but with, for now,
arbitrary cosmological constants).

\subsection{Coleman de Luccia}

The `CDL' bubble wall satisfies \eqref{rdot}, \eqref{tauplusdot},
and \eqref{tauminusdot}, which are solved by
\be
\beal
{\tilde R} &= \cos {\tilde\lambda}\\
{\tilde t}_- &=\frac{\alpha}{\sqrt{\alpha^2-1}} {\rm arctan}
\sqrt{\alpha^2-1} \sin{\tilde\lambda} \quad;\\
{\tilde t}_+ &=\frac{\alpha}{\sqrt{\alpha^2-(1-2{\bar\sigma}\gamma)^2}} 
{\rm arctan}\frac{\sqrt{\alpha^2-(1-2{\bar\sigma}\gamma)^2}}
{(1-2{\bar\sigma}\gamma)} \sin{\tilde\lambda} 
\eeal
\ee
(where ${\tilde\lambda}\in [-\frac\pi2,\frac\pi2]$ for the full bounce)
and the action can be computed analytically as
\be
B_{CDL} = - \frac{1}{2G} \int R({\dot\tau}_+ - {\dot\tau}_-)
= \frac{\pi}{G} \frac{{\bar\sigma}\gamma^3}
{\alpha(\alpha+1)(\alpha+1-2{\bar\sigma}\gamma)} 
\xrightarrow{\Lambda_+=0}\frac{\pi\ell^2}{G} \frac{16({\bar\sigma}\ell)^4}
{(1-4{\bar\sigma}^2\ell^2)^2}
\label{Bcdl}
\ee
Note that by analytic continuation, these expressions include 
arbitrary $\Lambda_\pm$, for which $\alpha<1$ or 
$1-2{\bar\sigma}\gamma$ are possible. In this special case 
the symmetry of the bubble solution has been raised from $O(3)$
to $O(4)$, and the result for the tunnelling rate agrees with 
explicitly $O(4)$ symmetric methods.  

\subsection{The general instanton}

The general bubble wall will have a black hole mass term on 
each side, and a general instanton will consist of a bubble 
trajectory between a minimum and maximum value of 
$\tilde R$. For fixed seed mass, $M_+$, there will be a range
of allowed $k_1$ and $k_2$ (see \eqref{physparams}), and
a corresponding range of values for the bounce action. 
By exploring the $\{k_1,k_2\}$ parameter space numerically 
and plotting the ratio of the bounce action to the CDL action,
we can build up a qualitative understanding of the preferred
instanton for vacuum decay. 

For example, if $\Lambda_+=0$, $GM_+ = \gamma k_1/2\alpha$,
and $GM_- = GM_+ - \gamma k_2 (1-\alpha)/\alpha^2$. 
Referring to figure \ref{fig:k1range}, we see there are two 
possibilities for the range of $k_2$, which is now a horizontal
line in the $k_1$ plot: Either the maximal 
value of $k_2$ lies on the $k_1^m$ branch with $GM_-=0$,
or on the static branch $k_1^*(k_2)$. The picture is
similar for general $\Lambda_+$, but the constant $GM_+$
lines are now at an angle, and interpolate between the $k_1^m$ 
curve at negative $k_2$ and either the $k_1^m$ line at positive 
$k_2$ or the $k_1^*(k_2)$ curve. The crossover
between the two possibilities occurs at $M_+=M_C$,
given by the algebraic solution to
\be
k_1^*(k_2) = \frac{2k_2}{\alpha} (1-\alpha)
\label{critk1k2}
\ee
when we have a static bubble with $GM_-=0$.
In either case, as $k_2$ drops, $GM_-$ increases until the 
lower limit of $k_2$ is reached at negative $k_2$ on the 
$k_1^m(k_2)$ curve. By solving 
numerically for the wall trajectories we find that the action
{\it increases} as $k_2$ drops. The preferred instanton therefore
is the one with the maximally allowed value of $k_2$ 
consistent with the value of $GM_+$. 

\begin{figure}[htb]
\centering
\includegraphics[width=0.7\textwidth]{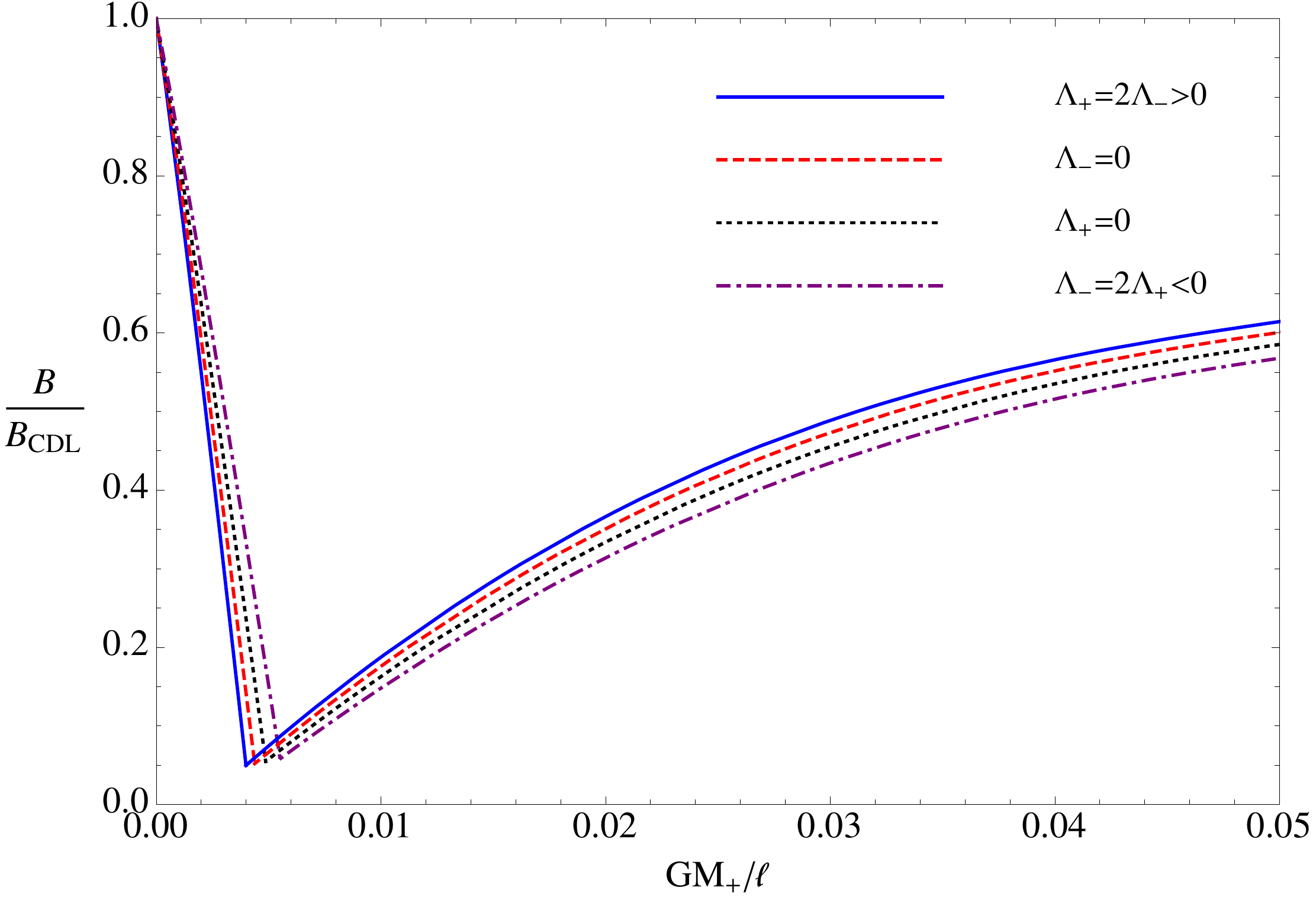}
\caption{A plot of the minimum bounce action as $M_+$ is
varied for ${\bar\sigma}\ell=0.1$, and varying values of
$\Lambda_+= 6/\ell^2, 3/\ell^2, 0, -3/\ell^2$, 
$\Lambda_-= 3/\ell^2, 0, -3/\ell^2, -6/\ell^2$ as indicated. The ratio of
the bounce action to the CDL value is plotted, but as 
$\Lambda_\pm$ vary, this value itself changes. For
${\bar\sigma}\ell=0.1$, 
$B_{CDL} = 0.101, 0.117, 0.137, 0.165\, \ell^2/L_p^2$
as $\Lambda_+$ drops from its maximal to minimal
value considered here.
}
\label{fig:lamcomp}
\end{figure}

This qualitative picture remains true irrespective of the 
values of $\Lambda_\pm$: for seed mass $M_+<M_C$, 
the dominant tunnelling process leaves behind a true
vacuum region and removes the black hole. The tunnelling 
rate is always faster than the vacuum tunnelling rate for these
instantons. The bounce action reaches a minimum at
$M=M_C$, where the bubble is static. For $M>M_C$ the 
dominant tunnelling process is a static bubble with a remnant 
black hole being left behind. As the seed mass increases further, 
eventually the tunnelling rate becomes lower than the vacuum 
tunnelling rate. Exploring the instantons for general $\Lambda$'s, 
we find that the ratio of $B/B_{CDL}$ changes very
little as the $\Lambda$'s vary. In figure \ref{fig:lamcomp},
we show how this dominant tunneling action varies 
as the values for the cosmological constants are changed. 
Since the change in $B/B_{CDL}$ is minimal (and $B_{CDL}$
itself is not varying much), we now restrict our discussion to
the $\Lambda_+=0$ set-up where $\alpha=1-2{\bar\sigma}\gamma$,
and many of the formulae simplify.

Before discussing the general dominant tunneling process,
we begin by considering the critical instanton where the
static bubble tunnels and removes the seed black hole 
altogether. Although \eqref{critk1k2} in general is a complicated
algebraic equation, for small ${\bar\sigma}\ell$ the various 
parameters can be expanded straightforwardly to give
\be
k_{1C} \simeq \frac{64}{27} ({\bar\sigma}\ell)^2
= \frac49 - 3k_{2C} \quad \Rightarrow \quad
\frac{GM_C}{\ell}\simeq \frac{128}{27} ({\bar\sigma}\ell)^3 
\label{Mcrit}
\ee
From \eqref{bounceaction}, the action of a static bounce in
general is
\be
B= {{\cal A}_+\over 4G}-{{\cal A}_-\over 4G} 
=4\pi GM_+^2-\pi G \left ( \frac{\ell}{G}\right)^{4/3}
({\mu}_+^{1/3}-
{\mu}_-^{1/3})^2,
\ee
where
\begin{equation}
G\mu_\pm=\sqrt{G^2M_-^2 + \frac{\ell^2}{27}}\pm 
\frac{GM_-}{\ell}
\end{equation}
although it must be noted that, for the static bubble $M_-$
is a complicated function of $M_+$. For the critical bubble,
$GM_-=0$, hence the critical bounce action is
\begin{equation}
B_C= 4\pi GM_C^2\simeq \frac{\pi \ell^2}{G} \left ( \frac{256}{27}\right)^2
({\bar\sigma}\ell)^6\,\simeq \left ( \frac{4}{3}\right)^6
({\bar\sigma}\ell)^2 B_{CDL}\,.
\end{equation}
Thus as ${\bar\sigma}\ell\to0$, the tunnelling action
becomes small compared to the CDL action.

One problem with having a small critical mass is of course
that the decay rate due to tunnelling may be outstripped
by the evaporation rate of the black hole, as we will discuss
later, however, what this expansion indicates is that the
minimal bounce action for a particular ${\bar\sigma}\ell$
can be extremely small, so that even if we are above
the critical black hole mass, the decay rate can still be
significant.

\begin{figure}[htb]
\centering
\includegraphics[width=0.7\textwidth]{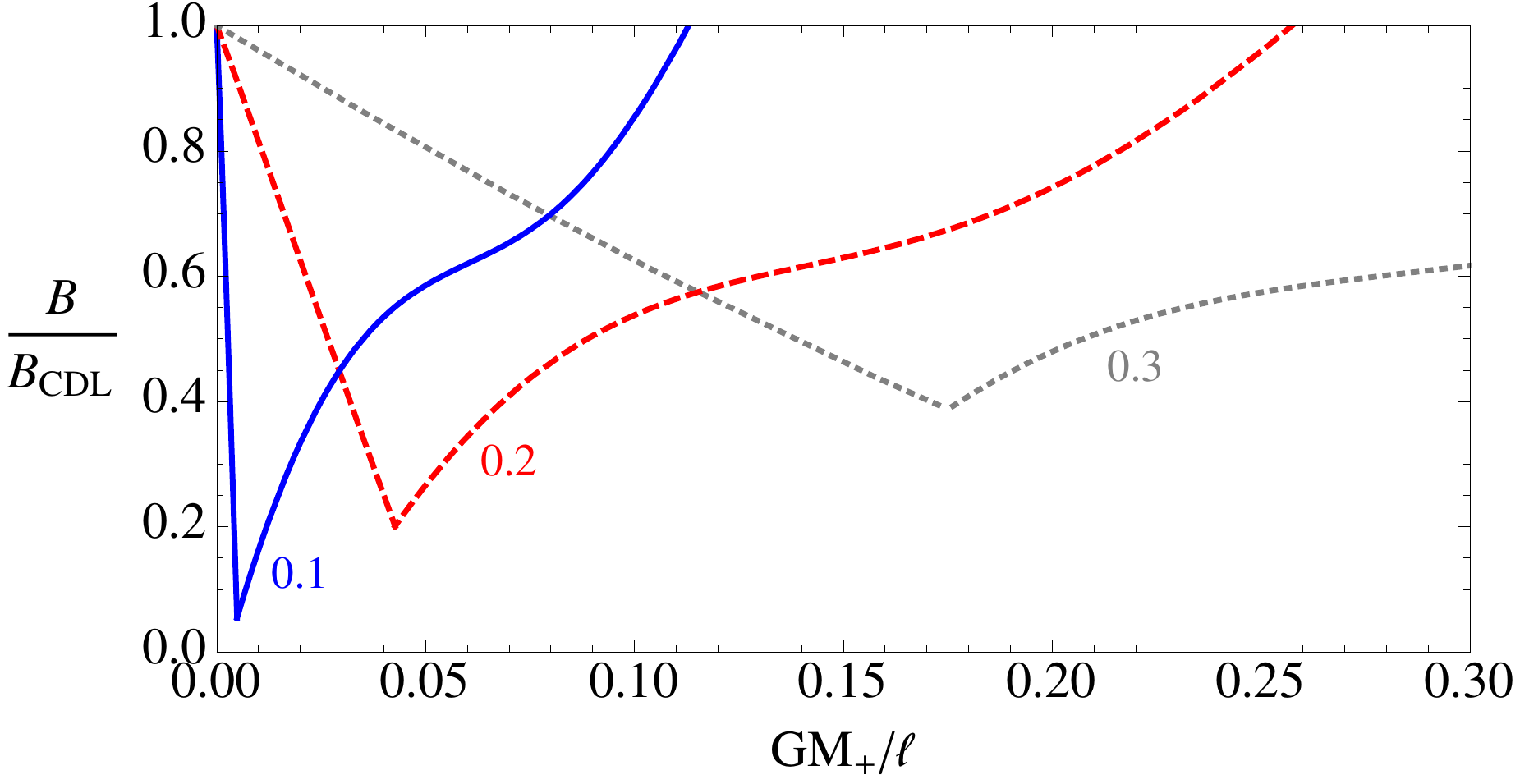}
\caption{
The exponent $B$ for the dominant tunnelling process 
divided by the appropriate vacuum tunnelling value $B_{CDL}$, for 
different masses $M_+$ of the nucleation seed. The surface tension 
$\sigma$ and AdS radius $\ell$ enter in the combination $\bar\sigma\ell$.}
\label{fig:domB}
\end{figure}

Having determined that the dominant tunneling process
is either the static bubble or the $GM_-=0$ branch, we
can now compute the dominant bounce action either
by numerically
solving the time-dependent bubbles with $GM_-=0$, or
computing the static bubble actions with $k_1=k_1^*$.
We used a simple mathematica program to calculate these
exponents, and double checked by a totally numerical computation.
The results for some sample values of
${\bar\sigma}\ell$ are presented in figure \ref{fig:domB}.

The general bubble solution for $GM_-=0$ oscillates between
two values ${\tilde R}_{MAX}$ and ${\tilde R}_{MIN}$ where 
the potential $U(\tilde R)$ vanishes. This periodic solution 
in $\tilde\lambda$ can only be single-valued in ${\cal M}_\pm$
if the manifolds on each side have the same time-periodicity 
as the bubble wall solution.
In general, this will not be the same as the natural periodicity 
$\Delta\tau_+=8\pi GM_+$ of the Euclidean Schwarzschild
solution, hence the need to consider general periodicity in 
the computation of the action in the previous section.
For the static solution of course, this is not an issue.
The values of ${\tilde R}_{MAX}$/${\tilde R}_{MIN}$ are
well outside the black hole horizon radius, and move together
as $GM_+$ is increased. Eventually, at $GM_C$, the two
roots of $U$ meet, and the static branch begins.

\begin{figure}[htb]
\centering
\includegraphics[width=0.8\textwidth]{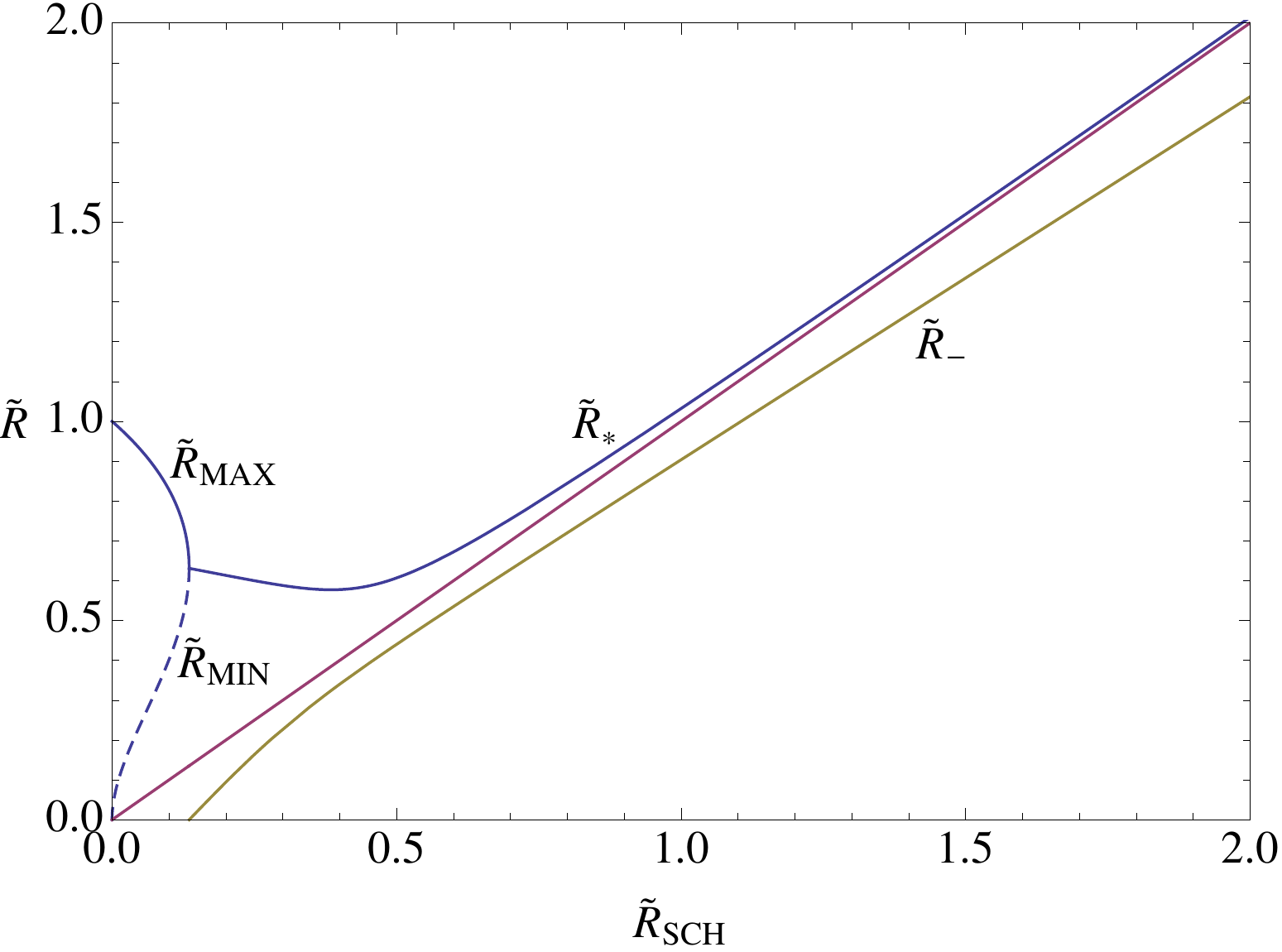}
\caption{
A plot of the variation of the bubble wall radius ${\tilde R}$ as 
$GM_+$ is increased for $\bar\sigma\ell=0.25$ (chosen to highlight
the qualitative features). As $\bar\sigma\ell$ drops, the features of 
the phase diagram remain the same, but `squash up' towards 
smaller ${\tilde R}_{SCH}$. The unlabelled red line running from
corner to corner represents 
${\tilde R}_{SCH}$, the seed black hole horizon radius.}
\label{fig:radii}
\end{figure}

The static branch is the preferred instanton with nonzero
$GM_-$, i.e.\ with a black hole remnant, although non-static
solutions exist with higher action and remnant mass.
Initially, the static bubble shrinks with increasing $GM_+$, but
remains well outside the Schwarzschild radius, however, as
we increase $GM_+$ further, the bubble becomes constrained by
the expanding black hole horizon, and becomes stretched just
outside ${\tilde R}_{SCH}$. Meanwhile, the remnant black hole mass
$GM_-$ increases along the static branch and eventually becomes
larger than $GM_+$, however, because of the negative cosmological
constant, the horizon radius, while increasing, does not increase 
as rapidly as ${\tilde R}_{SCH}$. The static bubble action therefore
increases as $GM_+$ increases, eventually becoming larger
than $B_{CDL}$ (see figure \ref{fig:domB}). Figure \ref{fig:radii}
illustrates the behaviour of these minimal/maximal and static
values of ${\tilde R}$ as ${\tilde R}_{SCH}=2GM_+$ varies, the
remnant horizon radius is also shown.

\subsection{Charged black hole instantons}

Finally, before considering the case of the Higgs vacuum in
detail, we conclude this section by commenting briefly on
an obvious generalisation of our instantons to Einstein-Yang-Mills-Higgs
theory.
The combination of Einstein gravity with Yang-Mills and Higgs fields 
admits the possibility of charged black hole solutions 
\cite{PhysRevD.11.2692,PhysRevD.12.1588}. Electrically charged
black holes can discharge by the emission of charged particles
\cite{Gibbons:1975kk}, but magnetically charged black holes may 
be the lightest magnetically charged particles in the theory, in which 
case a large mass black hole evaporates 
towards the extremal limit, and the Hawking radiation flux falls to zero.

Magnetically charged black holes may be produced in the 
early universe \cite{Mellor:1989gi,Mellor:1989wc},
and form the seeds for vacuum decay of an unstable standard model  
Higgs field. Uncharged black holes
can easily evaporate before they seed a phase transition, but the charged
black holes hang around for a longer time making them better candidates for
vacuum decay nucleation sites.

An $SU(2)\times U(1)$ Yang-Mills theory with Higgs field ${\cal H}$ 
in the fundamental $SU(2)$ representation has no flat-space monopole 
solutions, but it does have Dirac and Yang-Mills black-hole monoples. 
The non-abelian monopoles can be constructed from the $SU(2)$ fields 
$W$ using an ansatz
\begin{eqnarray}
{\cal H}&=&\phi(r)\sigma_r{\cal H}_0,\\
W&=&G^{1/2}{P\over r}
\left(\sigma_\phi\,d\theta-\sigma_\theta\sin\theta\,d\phi\right),
\end{eqnarray}
where $\sigma_r$, $\sigma_\theta$ and $\sigma_\phi$ are Pauli 
matrices projected along the spherical polar co-ordinate frame and 
${\cal H}_0$ is constant. (The magnetic charge has been scaled so 
that an extreme black hole has $P=M$ in the absence of a cosmological
constant.)

For a potential which allows decay from flat space to AdS, there
are thin-wall bubble solutions with spherical symmetry and 
constant values of $\phi$ at the appropriate minima of the potential. 
The metric coefficients are
\begin{eqnarray}
f_-&=&1-{2GM_-\over r}+{r^2\over \ell^2}+{G^2 P^2\over r^2},\\
f_+&=&1-{2GM_+\over r}+{G^2 P^2\over r^2}
\end{eqnarray}
In this case, the bubble wall carries no magnetic charge. Generalised 
solutions may also be possible in which the interior and exterior have 
different magnetic charges.

The action for the bubble solutions is given by the same formula,
\eqref{bounceaction}, as in the uncharged case, though with 
the appropriate expressions for $f_\pm$. The plot of the
dependence of the action on $GM_+/\ell$ is surprisingly similar
to the uncharged case at fixed ratio $P/M_+$, with one
small modification. The time-dependent tunneling solutions
prior to the switching on of the static bubbles now do not
remove the black hole altogether as this would leave a
naked singularity. Instead, the bubbles leave behind an
extremal remnant, $M_-=M_{ext}(P)$, where 
\be
GM_{ext}(P) = \frac{\ell}{3\sqrt{6}}
\left ( 2 + \sqrt{1+\frac{12G^2P^2}{\ell^2}}\right)
\sqrt{\sqrt{1+\frac{12G^2P^2}{\ell^2}} -1}\,.
\ee
The static branch meets this time-dependent branch
at a critical mass $M_{CP}$, where the static bubble
now has an extremal black hole in its interior.
On the static branch, the action is, as before, the difference
of the areas of the seed and remnant black holes, but as
the extremal limit is approached, the horizon radius of the
remnant black hole shrinks only as the root of $M_+-M_{CP}$,
whereas the radius of the seed black hole (which is not 
approaching an extremal limit) depends linearly on 
$M_+-M_{CP}$, thus, as we increase $M_+$ from 
$M_{CP}$, the action actually starts to {\it drop}
briefly, before the effect of the increasing horizon area
kicks in causing the usual rising of the bounce action.
This small dip in the action near the critical point is
very hard to see at low $P/M_+$, but for larger ratios
becomes more visible. The dip is however very minor,
and the minimum action is well approximated by
the value at $M_{CP}$. 
\begin{figure}[htb]
\centering
\includegraphics[width=0.8\textwidth]{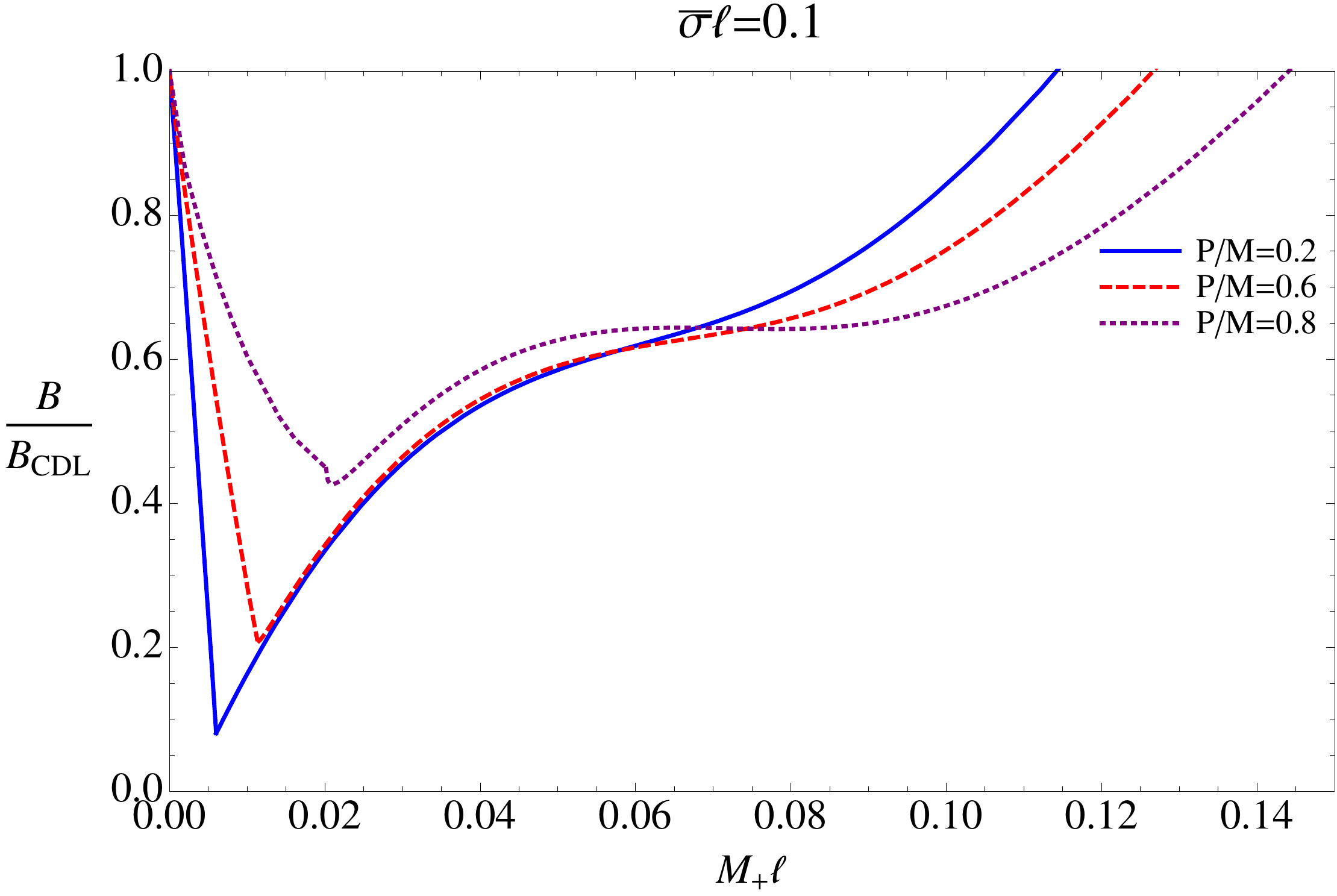}
\caption{
The exponent $B$ for the dominant tunnelling process with 
black-hole monopoles of mass
$M_+$ acting as nucleation seeds.}
\label{fig:BPplot}
\end{figure}

From figure \ref{fig:BPplot} we see the dip is most visible at
large ratio $P/M$, however, perhaps surprisingly, it is also 
the case that at large $P/M$ the catalytic effect of the 
black hole is much reduced. We therefore expect that
the addition of a monopole charge will not particularly 
assist with vacuum decay, a conclusion largely borne out
by the more detailed analysis of the next section.

\section{Application to the Higgs vacuum.}
\label{higgssec}

Up to this point, the vacuum decay process has been described 
in gravitational terms using the surface tension of the wall, 
$\sigma$, and the AdS radius of the `true' vacuum, $\ell$. 
In this section we will explore vacuum decay in the Higgs model 
with high energy corrections as discussed in the introduction. 
The key features of the potential relevant for quantum 
tunnelling are the barrier height, the separation between the 
minima and the energy of the true vacuum (TV).  These three 
parameters can be encoded as follows,
\begin{equation}
g=\phi_{TV}/M_p,\qquad 
\epsilon=-V(\phi_{TV}),\qquad 
\zeta=\sup\limits_{0<\phi<\phi_{TV}}\ 
V(\phi)
\end{equation} 

Following our previous discussion we shall restrict attention 
to potentials which allow thin-wall bubbles.
Although we would expect $\zeta\gg\epsilon$ for a thin 
wall bubble, numerical solutions show that the wall 
approximation is reasonably accurate even when 
$\zeta\sim\epsilon$, therefore we use this lower
bound for $\zeta$.
The range of Higgs model parameters $\lambda_*$, $\phi_*$ 
and $\lambda_6$ which allow thin-wall bubbles is set 
by $\zeta>\epsilon>0$, and by the condition that the true 
vacuum lies at large $\phi$. Thin wall bubbles correspond 
to rather large values of $\lambda_6$, as 
illustrated in table \ref{rates}. Roughly speaking, as
$\lambda_*$ becomes more negative, the values of 
$\lambda_6$ required for thin-wall become larger, similarly
as $\phi_*$ drops, $\lambda_6$ increases. In all cases
the pure CDL tunneling action is extremely large ($10^{6-7}$),
but the suppression of the critical tunneling action is also
large, and increases as $\lambda_*$ becomes more negative.

\begin{table}[ht]
\caption{A selection of parameter values for the modified Higgs potential, including
the AdS radius $\ell$, the rescaled surface tension $\bar\sigma\ell$ and
the critical mass $M_C$ for optimal nucleation seeded by a 
black hole. These parameters lie along the bottom edge of the parameter
ranges for thin-wall bubbles. The vacuum tunnelling exponent 
$B_{CDL}$ is around $4$e+06 in each of these examples.}
\vskip 1mm
\centering
\begin{tabular}{c c c c c c c c c c}
\hline\hline
$\lambda_*$ & $\phi_*/M_p$ &$\lambda_6$ &  $g$
&$\ell/L_p$&$\bar\sigma\ell$&$M_C/M_p$&$B_C/B_{CDL}$\\ [0.5ex]
\hline
-0.005 & 1  & 500 & 0.00146 & 3.17e+8 & 0.00045 & 3.5 & 1.2e-6\\
-0.005 & 0.5 & 2e+03&0.00073&   1.27e+9 & 0.00023 & 1.8 & 2.9e-7\\
-0.007 & 2 & 1.98e+03 & 0.0008 & 9.33e+8 & 0.00024 & 1.5 & 3.2e-7\\
-0.007 & 1 & 7.93e+03 & 0.0004 & 3.79e+9 & 0.00012 & 0.8 & 8.2e-8\\
-0.007 & 0.75 & 1.41e+04 & 0.0003 & 6.76e+9 & 9.1e-05 & 0.61 & 4.7e-8\\
-0.007 & 0.5 & 3.17e+04 & 0.0002 & 1.51e+10 & 6e-05 & 0.39 & 2e-8\\
-0.008 & 1 & 27e+03 & 0.00022 & 1.18e+10 & 6.9e-05 & 0.46 & 2.7e-8\\
-0.008 & 3 & 3e+03 & 0.00067 & 1.31e+9 & 0.00021 & 1.4 & 2.4e-7\\
-0.009 & 1 & 85e+03 & 0.00013 & 3.43e+10 & 4.1e-05 & 0.28 & 9.4e-9\\
-0.01  & 2 & 63e+03 & 0.00016 & 2.55e+10 & 5.4e-05 & 0.49 & 1.7e-8\\
\hline
\end{tabular}
\label{rates}
\end{table}

Following Coleman and De Luccia, we can express the surface 
tension of the bubble wall in terms of the potential. In order to 
extend the result to moderate values of $\zeta/\epsilon$,
we compute the tension using the integral 
\begin{equation}
\sigma=\int_0^{\phi_1} d\phi\,(2V)^{1/2}
\simeq \kappa gM_p\,\zeta^{1/2},
\end{equation}
where the upper limit of the integral is at $V(\phi_1)=0$. The 
constant $\kappa$ depends on the details of the potential, but
since $\phi_1<gM_p$ and $V\le\xi$, it is subject to the constraint 
$\kappa<\sqrt{2}$.
The AdS radius $\ell$ is related to the vacuum energy density by
\begin{equation}
\ell=\sqrt{3}M_p\epsilon^{-1/2}.
\end{equation}
The back-reaction parameter $\bar\sigma\ell$ is therefore
\begin{equation}
\bar\sigma \ell=\frac{1}{4 M_p^2}\sigma\ell=
\frac{\sqrt{3}}{4}\kappa g\left(\frac{\zeta}{\epsilon}\right)^{1/2}.
\end{equation}
Note that $\bar\sigma\ell<1/2$ puts an upper bound on $g$.
The CDL tunnelling exponent $B_{CDL}$ given in 
\eqref{Bcdl} is
\begin{equation}
B_{CDL} = 
\frac{27\pi^2\sigma^4}{2\epsilon^3(1-4\bar\sigma^2\ell^2)^2}
=\frac{27\kappa^4\pi^2}{2}
\left(\frac{g^4M_p^4}{\epsilon}\right)
\left(\frac{\zeta}{\epsilon}\right)^2
(1-4\bar\sigma^2\ell^2)^{-2}.
\end{equation}
The large size of $B_{\rm CDL}$ in the parameter range covered 
by table \ref{rates} guarantees a tunnelling lifetime longer than 
the age of the universe (for unseeded nucleation).

When the vacuum decay is seeded by a black hole, 
the most rapid decay process occurs
for a seed mass $M_+=M_C$ given in \eqref{Mcrit},
\begin{equation}
M_C
\approx{128\over 27}{\ell\over G}(\bar\sigma\ell)^3
=\frac{16}{3}\pi\kappa^3\,
\left({g^4M_p^4\over\epsilon}\right)^{1/2}
\left({\zeta\over\epsilon}\right)^{3/2}
M_p.
\label{mcc}
\end{equation}
The corresponding exponent in the nucleation rate is 
$B_C=0.5(M_C/M_p)^2$.
Some values for $M_C$ are shown in table  \ref{rates}.
If $M_C\gg M_p$, then the exponent $B_C$ is large and 
the seeded decay rate becomes vanishingly small. On the 
other hand, if $M_C \lesssim M_p$ then even if we have a
seed mass $M_+ \gg M_C$, we can still get significant
suppression of the bounce action while remaining well above
the Planck mass from tunneling on the static branch,
as we see from figure \ref{fig:domB}. Our strategy therefore
is to explore the decay seeded by a small mass black hole
via the static instanton, which (for convenience) we 
determine numerically.
\begin{figure}[htb]
\centering
\includegraphics[width=0.7\textwidth]{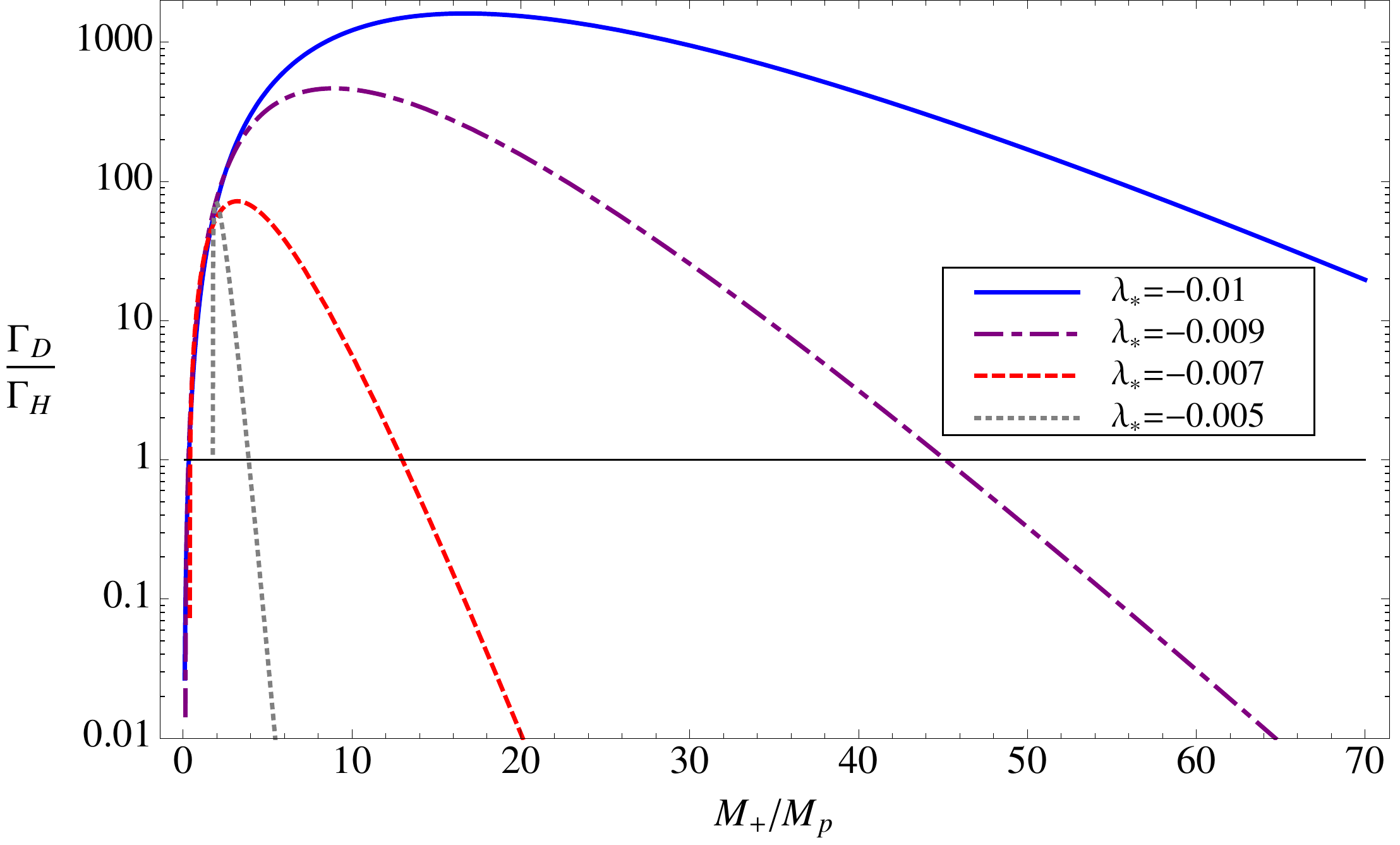}
\includegraphics[width=0.7\textwidth]{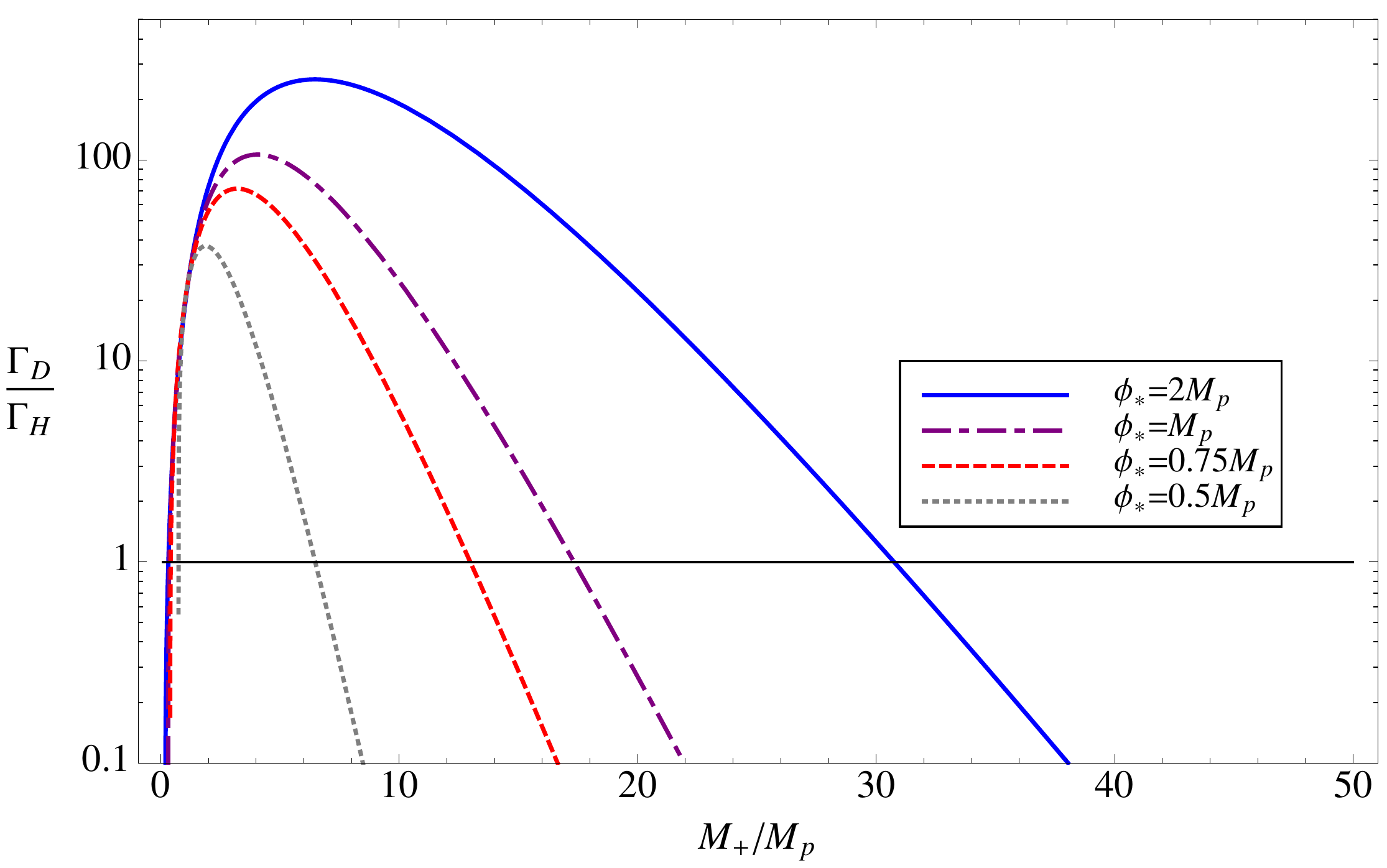}
\caption{
The branching ratio of the false vacuum nucleation rate to 
the Hawking evaporation rate as a function 
of the seed mass for a selection of Higgs models from table 
\ref{rates}. The first plot shows the branching ratio for
$\phi_*=M_p$ with the labelled values of $\lambda_*$, and
the second for $\lambda_*=-0.007$ for the labelled values
of $\phi_*$.
The black hole starts out with a mass beyond the right-hand 
side of the plot and the mass decreases by Hawking evaporation. 
At some point, the vacuum decay rate becomes larger
than the Hawking evaporation rate.}
\label{fig:ratio}
\end{figure}

A brief consideration of the dependence of the bounce action 
on $M_+$ shows that we are exploring seeded tunnelling for very
light or primordial black holes \cite{Carr:1974nx}, with temperatures 
well above that of the CMB. We must therefore check that 
the black hole can seed the false vacuum decay
before it disappears through Hawking radiation.
The vacuum decay rate $\Gamma_D$, \eqref{GammaD},
contains not only the exponential of the bounce action, but
also a pre-factor, $A$.  According to Callan 
and Coleman \cite{callan1977}, this pre-factor is made up 
of a factor of $(B/2\pi)^{1/2}$ for each translational zero 
mode of the instanton and a determinant factor. 
In our case, there will be a single zero mode representing 
the time translation symmetry, and rather than evaluate the 
determinant factor, we use the inverse horizon timescale 
as a rough estimate $(GM_+)^{-1}$,  giving
\begin{equation}
\Gamma_D\approx \sqrt{\frac{B}{2\pi}} \,
\frac{e^{-B}}{GM_+}.
\end{equation}
The black hole emits Hawking radiation at a rate depending 
on fundamental particle masses and spins. The total decay rate 
for a subset of the standard model was evaluated by Page, 
\cite{Page}. If we set $\Gamma_H=\dot M/M$, then
\begin{equation}
\Gamma_H\approx 3.6\times 10^{-4}(G^2 M_+^3)^{-1}
\end{equation}
The branching ratio of the tunnelling rate to the evaporation rate for 
uncharged black holes is therefore
\begin{equation}
\frac{\Gamma_D}{\Gamma_H}\approx
44{M_+^2\over M_p^2}B^{1/2}e^{-B}.
\end{equation}

From this expression we can see roughly how the branching 
ratio will depend on $M_+$, even though $B$ is, in principle,
a complex function of $M_+$. The static bubble is the difference
in areas of the seed and remnant black holes, which we can
deduce from figure \ref{fig:radii} to be roughly linear (there is actually
a slightly stronger dependence on $M_+$, however, what is
important is that it is not quadratic), whereas the prefactor is
(again, roughly) $M_+^{5/2}$; we therefore expect the plot to
be strongly exponentially suppressed at large $M_+$, but rising
as $M_+$ falls to a maximum around $M_+/M_p = {\cal O}(1)$,
then dropping again below $M_p$. The actual value of the
maximum will depend on the details of how $B$ depends on $M_+$,
which requires a full calculation.

The branching ratio is plotted as a function of the seed 
mass $M_+$ for two indicative sets of parameters taken from
table \ref{rates} in figure 
\ref{fig:ratio} in order to illustrate the dependence on the
parameters in the Higgs potential (or correspondingly on
${\bar \sigma}$ and $\ell$). The branching ratio is shown at
fixed $\phi_*$ with varying $\lambda_*$ and vice versa.
The overall picture is that for lower ${\bar\sigma}$ and 
higher $\ell$ (or more negative $\lambda_*$ / higher 
$\phi_*$) the branching ratio is larger, and is consistently
higher than unity over a larger range. While Hawking evaporation
always dominates at large $M_+$, the effect of Hawking radiation
is that the black hole loses mass, hence driving it towards 
increasing branching ratio. A black hole produced in the early 
universe, for example, starts out with a mass well beyond the 
right-hand side of the plots, but at some point after evaporation,
the vacuum decay rate becomes larger than the Hawking 
evaporation rate and the black hole seeds the transition to a 
new vacuum. This can occur for seed masses well above the 
Planck mass, where we have some confidence in the validity 
of the vacuum decay calculation. The timescales for Hawking 
evaporation and vacuum decay will both be less 
than roughly a million Planck times.

\begin{figure}[htb]
\centering
\includegraphics[width=0.7\textwidth]{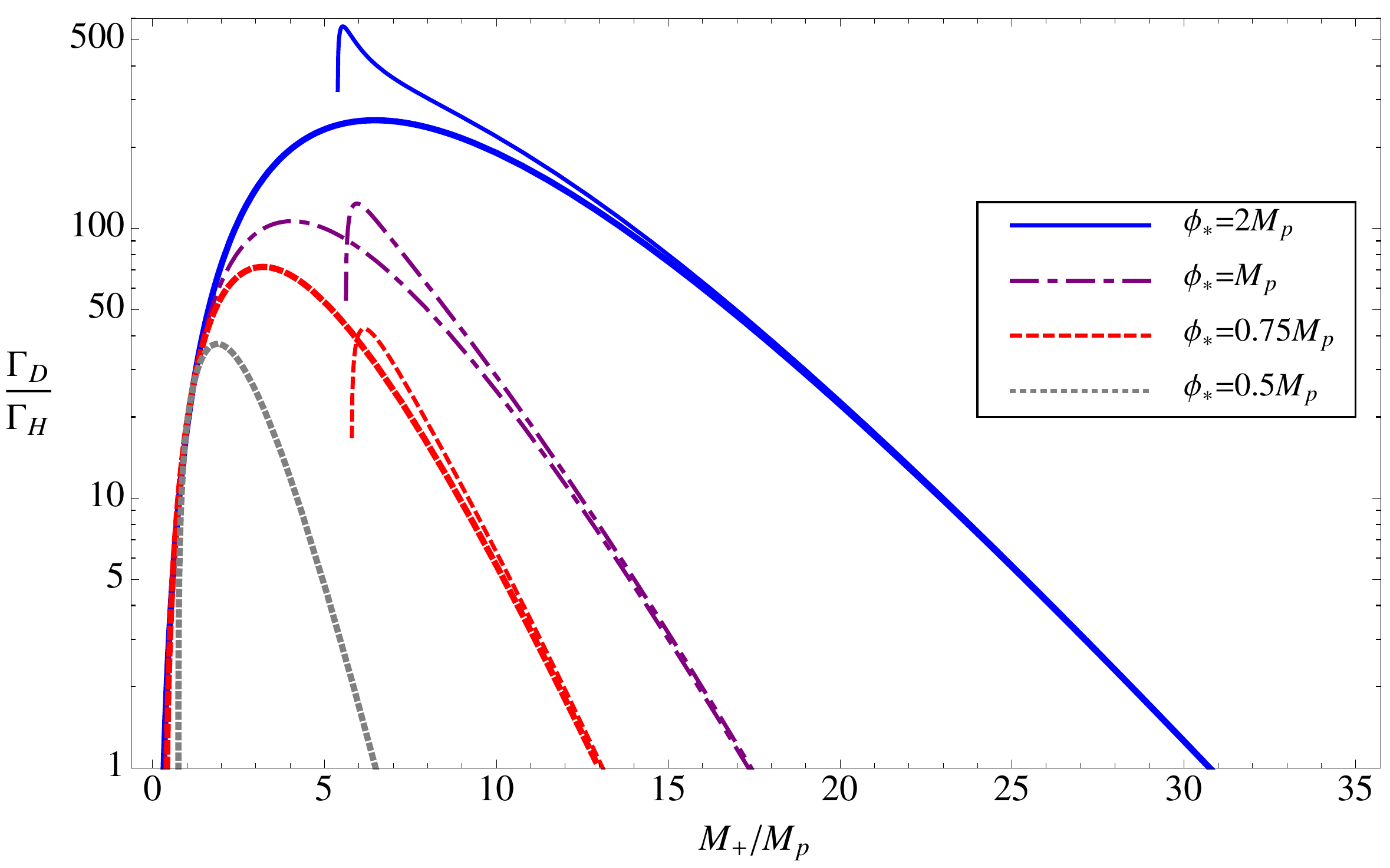}
\caption{
The branching ratio of the false vacuum nucleation rate to 
the Hawking evaporation rate for a monopole charged 
black hole with $P=5M_p$, shown as a function 
of the seed mass for a selection of Higgs models from table 
\ref{rates}. The plot for the uncharged black hole ($P=0$) 
is repeated for comparison. As before,
$\lambda_*=-0.007$ for the labelled values
of $\phi_*$.
}
\label{fig:ratiomon}
\end{figure}

Finally, for the case of a monopole charged black hole, we might
expect the branching ratio to be larger due to their reduced Hawking 
radiation rate: The Hawking flux is proportional to 
${\cal A}_+T_+^4$, where ${\cal A}_+$ is the event horizon, 
and $T_+$ is the Hawking temperature
\begin{equation}
T_+={1\over 8\pi G M_+}\frac{4\Delta}{(1+\Delta)^2},
\end{equation}
setting $\Delta^2=1-P^2/M_+^2$. The evaporation rate is now
\begin{equation}
\Gamma_H\approx 3.6\times 10^{-4}(G^2 M_+^3)^{-1}
\frac{64\Delta^4}{(1+\Delta)^6}
\end{equation}
The branching ratio $\Gamma_D/\Gamma_H$ can now be re-computed
using this evaporation rate and the vacuum decay rates from the previous
section. The result is shown
in figure \ref{fig:ratiomon}.

\section{Conclusions}
\label{discuss}

The main aim of this paper was to demonstrate that 
black holes can massively speed up the rate of decay of a 
metastable Higgs vacuum by acting as nucleation sites. 
We have shown that is the case, subject to the limitations of 
the analysis, some of which we shall now address.

We have used a Higgs model which was based loosely on the 
two-loop effective potential, with an extra term motivated by 
high-energy physics, possibly quantum gravity. This has enabled 
us to simplify the analysis by employing a thin-wall approximation, 
which is valid in part of the parameter space. The thin-wall 
approximation shows that when the seed mass is above some 
(calculable) critical mass, then the seeded nucleation 
proceeds via a static bubble solution.
This gives a good starting point for an analysis of the thick wall 
bubble nucleation, which is far simpler for static than for non-static 
field configurations. We will show in a companion paper that 
the thick-wall bubble solutions extend the allowed 
range of parameters to small values of our new coefficient 
$\lambda_6$, which was rather large here, 
to the limit where $\lambda_6=0$, and the potential becomes 
simply the two-loop Higgs potential 
from the standard model.

The seeded nucleation calculation presented here requires 
a black hole, and for large tunneling enhancement,
this is expected to be a primordial black hole which is 
evaporating and nearing the end of its life. There is of course
another situation in which small black hole might occur. A possible
alternative solution to the hierarchy problem has been to
consider large extra dimensions, 
\cite{ArkaniHamed:1998rs,Antoniadis:1998ig,Randall:1999ee,
Randall:1999vf}. In these models, our
universe is presumed to be a ``brane'' living within higher dimensions
on which standard model physics is confined. The higher dimensional
Planck scale is not hierarchically large, but instead our 4D Planck 
scale, which is derived via an integration over these extra dimensions,
gains its leverage via the ``large'' internal volume. In such scenarios
black holes can be created in particle collisions 
\cite{Giddings:2001bu,Dimopoulos:2001hw}, leading to
considerable interest in the possibility of black holes being
produced at the LHC (for a recent review see \cite{Kanti:2014vsa}).
There are no known exact solutions for these black hole plus 
brane systems, and instead the black hole is usually considered
to be approximately a higher dimensional Schwarzschild
or Myers-Perry \cite{Myers:1986un} solution (see
\cite{Kanti:2004nr,Gregory:2008rf} for reviews on the
issues and properties of brane black holes). For the
Randall-Sundrum braneworld models, where the extra
dimension is strongly warped, one could also consider
``brane only'' solutions, such as the tidal black hole
\cite{Dadhich:2000am}.

Calculating the vacuum decay rates for these systems would 
be challenging, to say the least, not only because of the lack of
a true higher dimensional black hole solution, but also because an
instanton presumably would have to have a different vacuum
only on the brane, and not in the bulk (although the 
braneworld equivalent of the CDL instantons were 
constructed in \cite{Gregory:2001xu,Gregory:2001dn}).
However, some 
features of our calculation should be present.
The tidal black holes, for example, resemble 
black-hole monopoles, but with negative
square monopole charge $P^2$. The bubble solutions for tidal
holes are simple generalisations of the ones we have already discussed. 

\begin{figure}[htb]
\centering
\includegraphics[width=0.8\textwidth]{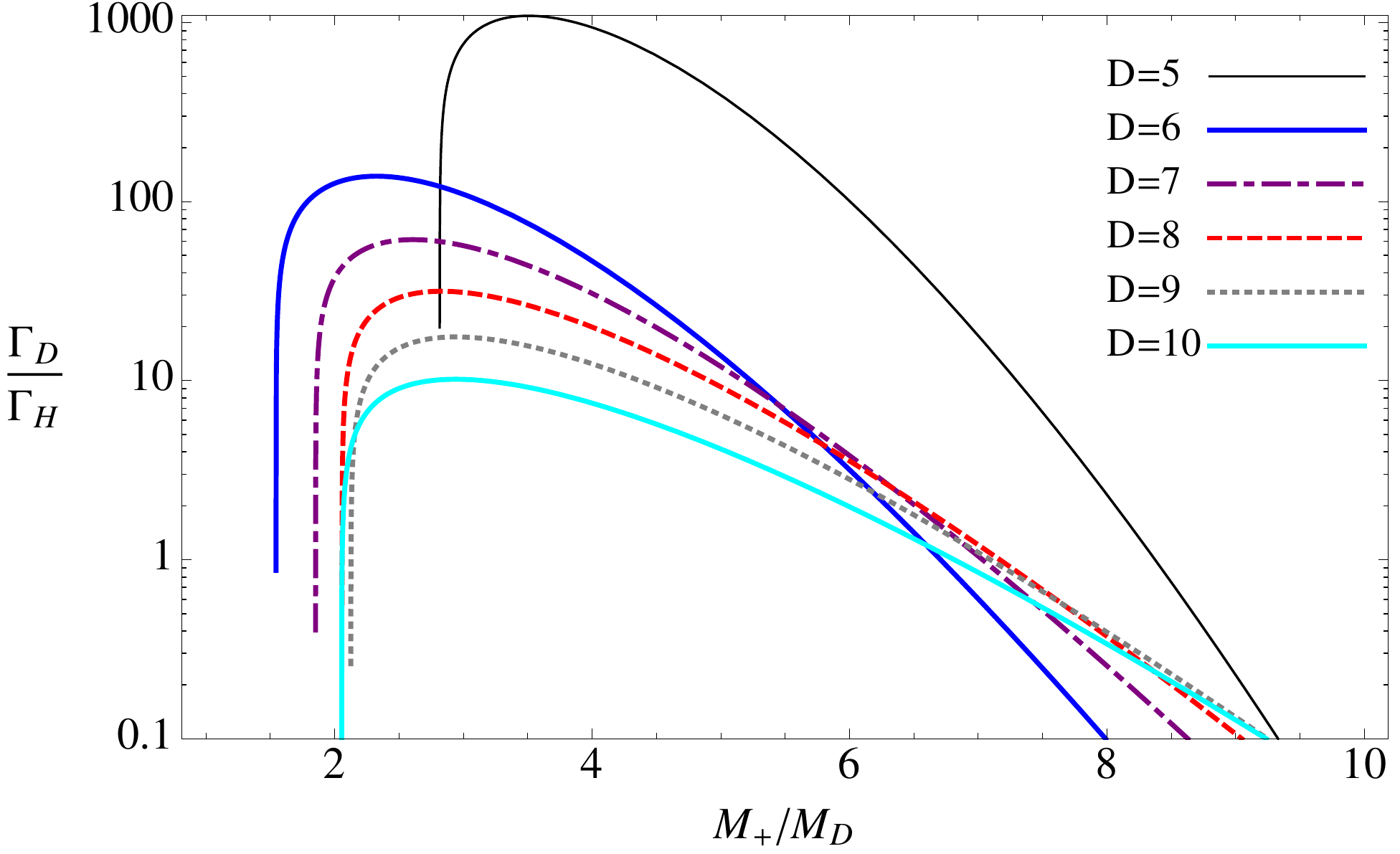}
\caption{
The dependence of the branching ratio on the dimensionality
of spacetime, $D$. Here, the $D-$dimensional Planck mass
is fixed, i.e.\ $8\pi G_D$ always has the same value.
}
\label{fig:branchD}
\end{figure}

More concretely, in analogy with the use of the Myers-Perry
geometries to approximate the small collider black holes,
we could consider a higher dimensional analogue
of our instanton. This would be a flawed model as it
does not fit easily with a Higgs field which is confined 
to the brane, but it provides a concrete calculational
method presumably as reasonable as the use of higher 
dimensional black hole geometries to approximate the
brane black hole. Appendix \ref{higherD} gives the details
of the computation with the higher dimensional instanton,
which leads to a dimension dependent branching ratio
\be
\frac{\Gamma_D}{\Gamma_H} \approx
550 \frac{M_+ r_+}{(D-3)^4} \sqrt{B_D} e^{-B_D}
= \frac{550 \sqrt{B_D} e^{-B_D}
}{(D-3)^4}\left [ \frac{16\pi G_DM_+^{D-2}}{(D-2)A_{D-2}}
\right]^{1/(D-3)}
\label{branchD}
\ee
where $B_D$ is given (implicitly) in \eqref{BDdef}.
Defining the higher dimensional reduced Planck mass
as\footnote{Note that in the literature, see e.g.\ \cite{Kanti:2004nr},
the non-reduced Planck mass is often used. Due to the
dimension dependence of the Planck mass this will 
introduce various dimension dependent renormalisation 
factors between our expressions and those assumed there.
Although these are of order unity, they do have some impact.}
\be
M_D^{D-2} = \frac1{8\pi G_D}
\ee
we can track the branching ratio as a function of $M_+/M_D$
and its dependence on $D$. To illustrate this dimensional
dependence, we chose test-case values
of ${\bar\sigma}\ell=0.01$ and $\ell=0.1$, and 
plotted $\Gamma_D/\Gamma_H$ in figure \ref{fig:branchD}.

The figure shows that as the number of extra dimensions increases,
the branching ratio decreases, however the exact normalisation of
the plot will depend on the confidence of translating modifications
of the Higgs potential to the new Planck scale and variables. Given
the crudeness of this particular model, we leave this, and possible
refinements of the decay modelling to future work.

In conclusion, we have shown that the lifetime of our universe 
in a metastable Higgs phase is crucially dependent on the absence 
of any nucleation seeds, and a primordial black hole
could drastically reduce the time it takes to decay onto a 
different `standard model'. 
Instability of the standard model is therefore more problematic 
than was hitherto supposed. Further exploration of the parameter 
space, using a wider class of bubble nucleation scenarios, should 
give us the range of Higgs parameters which lead to a long-lived
standard model in the presence of black holes.

\section*{Acknowledgements} 

We would like to thank Ben Withers for collaboration in the early stages 
of this project, and also Niayesh Afshordi for useful discussions.

PB is supported by an EPSRC International Doctoral Scholarship,
RG and IGM are supported in part by STFC (Consolidated Grant ST/J000426/1).
RG is also supported by the Wolfson Foundation and Royal Society, and
Perimeter Institute for Theoretical Physics. 
Research at Perimeter Institute is supported by the Government of
Canada through Industry Canada and by the Province of Ontario through the
Ministry of Research and Innovation. 

\appendix

\section{The limits on $k_1, k_2$}
\label{kappalimits}

The values of $k_1$ and $k_2$ are limited by requiring 
existence of a solution to ${\dot {\tilde R}}^2 + U({\tilde R})=0$,
where $U$ is defined in \eqref{rdot}, requiring positivity of
the black hole masses, and positivity of the arrows of time 
on each side of the wall:
\begin{eqnarray}
&&f_+ \frac{d{\tilde\tau}_+}{d{\tilde\lambda}}
= \frac{k_2}{{\tilde R}^2} + \frac{\tilde R}{\alpha} 
(1-2{\bar\sigma}\gamma)\label{tauplusdot}\\
&&f_- \frac{d{\tilde\tau}_-}{d{\tilde\lambda}}
= \frac{k_2}{{\tilde R}^2} + \frac{\tilde R}{\alpha} 
\label{tauminusdot}
\end{eqnarray}
where
\be
\beal
f_+ &= 1 - \frac{k_1}{\tilde R} - \frac{2k_2}{\alpha\tilde R} 
\left ( \alpha - (1-2{\bar\sigma}\gamma)\right)
- \frac{{\tilde R}^2}{\alpha^2}  
\left[ \alpha^2 - (1-2{\bar\sigma}\gamma)^2\right]\\
f_- &= 1 - \frac{k_1}{\tilde R} + \frac{2k_2}{\alpha\tilde R} (1-\alpha)
+ \frac{{\tilde R}^2}{\alpha^2}  (1-\alpha^2)\;.
\eeal
\ee
The first requirement is algebraically identical to the 
constraint discussed in \cite{GMW} (although the expression
given there in the appendix was not correct). Simultaneously requiring
$U=U'=0$ and eliminating ${\tilde R}_\ast$ gives an upper 
limit $k_1\leq k_1^*$, the correct expression for which is:
\be
\begin{aligned}
k_1^\ast &= \frac{2}{9} \Biggl [ 1 +81 k_2^2- \left ( 
-1 -5 (27k_2)^2 + \frac{(27k_2)^4}{2} + \frac{27k_2}{2} 
\left ( 4 + (27k_2)^2\right)^{3/2} \right)^{1/3}\\
& \qquad + \left ( 1 +5 (27k_2)^2 - \frac{(27k_2)^4}{2} 
+ \frac{27k_2}{2} \left ( 4 + (27k_2)^2\right)^{3/2} \right)^{1/3}
\Biggr]^{1/2} -2k_2
\end{aligned}
\label{kappa1star}
\ee

To get lower limits for $k_1$ we have to consider positivity of the black
hole masses and the arrows of time on each side of the wall. These now
depend on the sign and magnitude of the cosmological constants and
are different to \cite{GMW}. First, note the relation between the physical 
quantities and the parameters:
\be
\beal
GM_- &= \frac{\gamma}{2\alpha}\left ( k_1
- 2 \frac{(1-\alpha)}{\alpha} k_2 \right)\\
GM_+ &= \frac{\gamma k_1}{2\alpha}
+ \frac{k_2\gamma}{\alpha^2} \left(
\alpha-1+ 2{\bar\sigma}\gamma \right )\\
\Lambda_- &= 3 \frac{\alpha^2-1}{\gamma^2}\\
\Lambda_+ &= \Lambda_- -12 {\bar\sigma}^2 + 12\frac{\bar\sigma}{\gamma}
= \frac{3}{\gamma^2} \left ( \alpha^2-(1-2{\bar\sigma}\gamma)^2\right)
\eeal
\label{physparams}
\ee
thus
\be
\beal
GM_- \geq 0 \quad &\Rightarrow \qquad
k_1 \geq 2(1-\alpha) \frac{k_2}{\alpha}\\
GM_+ \geq 0 \quad &\Rightarrow \qquad
k_1 \geq 2(1-\alpha+2{\bar\sigma}\gamma) \frac{k_2}{\alpha}
\eeal
\label{posmasses}
\ee
Secondly, the requirement of positivity of ${\dot{\tilde\tau}}_\pm$, 
of which the constraint on $\dot{\tilde\tau}_+$ is the stronger:
\be
\frac{k_2}{{\tilde R}^2} + \frac{\tilde R}{\alpha} 
(1-2{\bar\sigma}\gamma) \geq 0
\ee
saturated by ${\tilde R}_+^3 = \alpha k_2 /(2{\bar\sigma}\gamma-1)$. 
Clearly, if $2{\bar\sigma}\gamma>1$, we must have $k_2>0$, and
$U$ must be positive with positive gradient at ${\tilde R}_+$. A brief
manipulation of $U'>0$ yields
\be
\frac{\alpha^2}{(2{\bar\sigma}\gamma-1)^2} -2
-\frac{k_1+2k_2}{2k_2} \frac{\alpha}{(2{\bar\sigma}\gamma-1)} \geq 0
\ee
From this, we see that $\Lambda_+>0$,
$2{\bar\sigma}\gamma \leq 1+\alpha/2$, and $k_1$ is bounded
below by $k_1^m=2 (1-\alpha)k_2/\alpha$ from \eqref{posmasses}.
\begin{figure}[htb]
\centering
\includegraphics[width=0.7\textwidth]{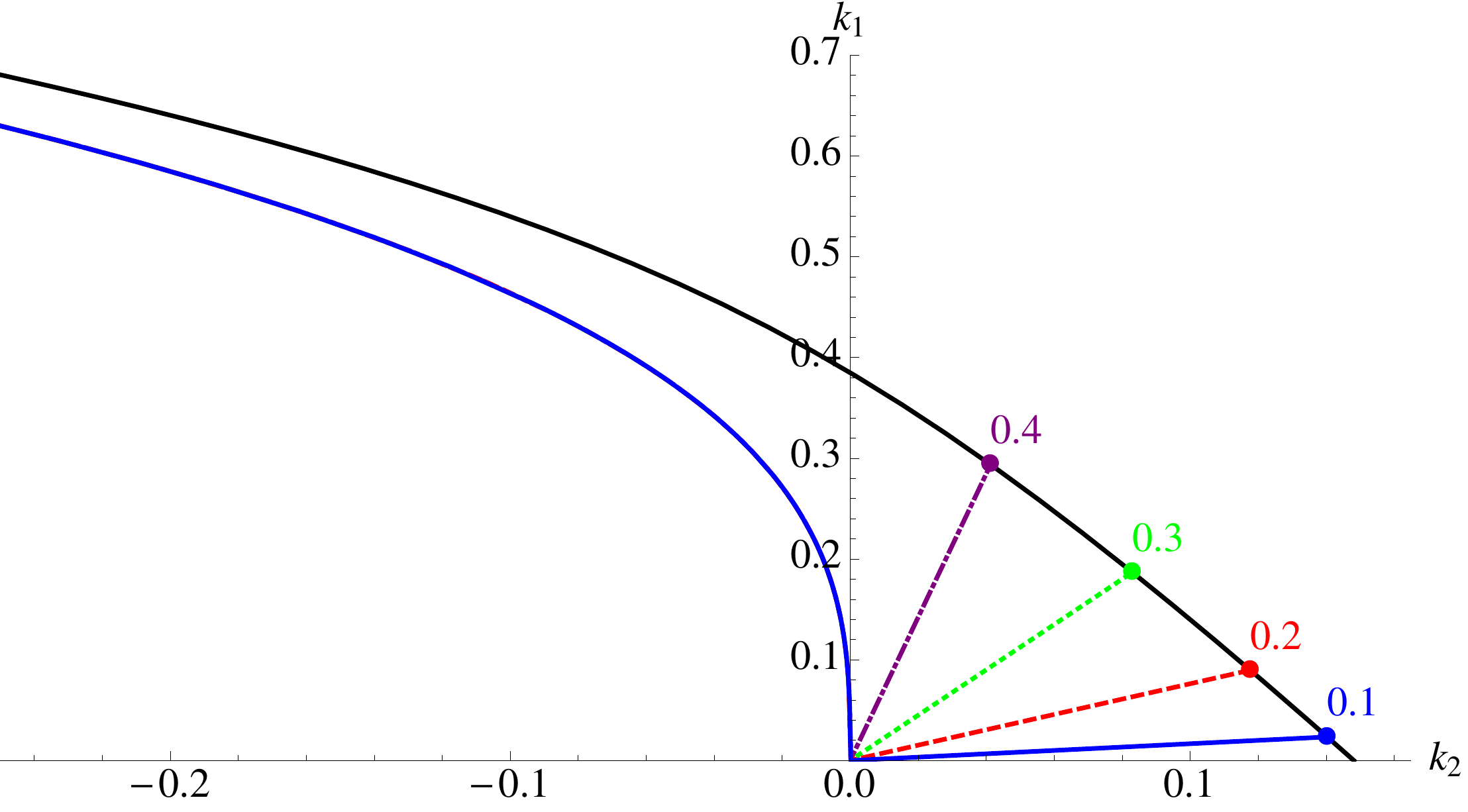}
\caption{
The allowed ranges for the parameters $k_1$ and $k_2$
with $\Lambda_+=0$. 
The upper bound on $k_1$ corresponds to stationary
wall solutions. The lower bound on $k_1$ when $k_2>0$ 
corresponds to vanishing remnant mass $M_-=0$. 
The two limits intersect at a point $(k_{1C},k_{2C})$, which depends on the
$\bar\sigma\ell$ and approaches $(0,4/27)$ as $\bar\sigma\ell\to 0$.
The strip of allowed $k_1$ continues to the left as $|k_2|^{1/3}$.}
\label{fig:k1range}
\end{figure}

Now consider $2{\bar\sigma}\gamma<1/2$, for which $k_2<0$ is
allowed. A similar argument to \cite{GMW} gives
\be
k_1\geq k_1^m=
\left | \frac{\alpha k_2}{2{\bar\sigma}\gamma-1}\right| ^{1/3} +
\frac{k_2}{(1-2{\bar\sigma}\gamma)} 
\frac{(\alpha-1+2{\bar\sigma}\gamma)^2}{\alpha}
\label{k1minnegk2}
\ee
for $k_2\leq0$, and once again, $k_1^m=2 (1-\alpha)k_2/\alpha$ 
for $k_2>0$. Note that neither of these bounds requires a particular
sign for $\Lambda_-$.

Where the sign of $\Lambda_-$ does make a difference is in
the range of allowed $k_2$. The upper limit on $k_2$ is determined
by $M_-=0$ for the static bounce, i.e.\ 
\be
\alpha k_1^*[k_2] = 2(1-\alpha)k_2
\ee
and a lower bound on $k_2$ is determined when the range of $k_1$
for $k_2<0$ closes off at negative $k_2$, which occurs when
$U({\tilde R}_+)=U'({\tilde R}_+)=0$:
\be
{\tilde R}_+^4-3{\tilde R}_+^6+3k_2^2 =
{\tilde R}_+^4 - \frac{\gamma^2 k_2^2}{(1-2{\bar\sigma}\gamma)^2}
\Lambda_+ = 0
\ee
which is clearly inconsistent for $\Lambda_+\leq0$.
For $\Lambda_+>0$, the range closes off at
\be
k_{2,min} = \frac{\alpha^2(2{\bar\sigma}\gamma-1)}
{3\sqrt{3}[\alpha^2 - (1-2{\bar\sigma}\gamma)^2]^{3/2}} 
 = \frac{\alpha^2(2{\bar\sigma}\gamma-1)}
{\gamma^3 \Lambda_+^{3/2}} 
\ee
Note that for $\Lambda_+<0$, the range of $k_1$, while
initially narrowing as $k_2$ becomes negative, eventually
opens out, as the linear term in \eqref{k1minnegk2} becomes
dominant. Thus large AdS black holes can tunnel to an even larger
AdS black hole with a more negative cosmological constant.

In this paper, we mostly focus on $\Lambda_+=0$, for which
$\alpha = 1-2{\bar\sigma}\gamma$, and many of the expressions
above simplify:
\be
k_1^m = 
\begin{cases}
\left (- k_2\right ) ^{1/3} & k_2\leq0\\
\frac{2(1-\alpha)}{\alpha}k_2 & k_2>0
\end{cases}
\ee
and the range of $k_2$ is plotted in figure \ref{fig:k1range}.
Note that unlike the figure in \cite{GMW}, where the lower limit
for $k_1$ was dependent on ${\bar\sigma}$ for negative but
not positive $k_2$, in this case it is the lower limits for positive
$k_2$ and not negative that depend on $\bar\sigma$.

\section{Alternative bounce action calculation}
\label{altaction}

In this appendix we present an alternative derivation of the
general expression for the bounce action using the Hamiltonian
approach presented in \cite{GMW}, and extend the result in 
some cases beyond the thin-wall limit. We will evaluate the 
Euclidean action for gravity plus a scalar field, with Lagrangian 
${\cal L}_m$, in an asymptotically AdS or flat spacetime. 
As in \textsection \ref{actioncomputation}, we take the action
on a sequence ${\cal M}_r$ of manifolds with boundary 
$\partial{\cal M}_r$ at $r_0$, and subtract the action of a 
similar sequence $\overline{\cal M}_r$ of Euclidean AdS manifolds 
with the same boundary $\partial{\cal M}_r$. 

It is instructive to first consider the case where the Euclidean 
spacetime $\cal M$ is perfectly regular, with no conical singularities,
and has a Killing vector, $\partial_\tau$. We perform a foliation of 
${\cal M}_r$ with a family of non-intersecting surfaces $\Sigma_\tau$ 
(assuming the global topology permits), with $0<\tau<\beta$. 
The canonical decomposition of such foliations has been investigated by 
Hawking and Horowitz, \cite{Hawking:1995fd}. 
In order to decompose the action, we use their identity
\begin{equation}
{\cal R}={}^3{\cal R}-K^2+K_{ab}^2-2\nabla_a(u^a\nabla_b u^b)
+2\nabla_b(u^a\nabla_a u^b), 
\label{curvatureidentity}
\end{equation}
where the vector $u^\mu$ is normal to $\Sigma_\tau$ and ${}^3{\cal R}$
is the Ricci-curvature of $\Sigma_\tau$.
The action therefore splits into bulk and boundary parts,
\begin{align}
I_{\mathcal{M}_r}=&-\frac{1}{16\pi G}\int_0^\beta\,d\tau
\int_{\Sigma_\tau} \left({}^3{\cal R}-K^2+K_{ab}^2
+16\pi G{\cal L}_m\right)\sqrt{g}\nonumber\\
&\qquad +\frac{1}{8\pi G}
\int_{\partial{\cal M}_r}n_b u^a\nabla_a u^b\sqrt{h}.
\label{Ifull}
\end{align}
The bulk term expressed in terms of canonical momenta 
$\pi^{ij}$ and $\pi$ becomes
\begin{equation}
\frac{1}{16\pi G}\int_0^\beta\,d\tau\left[
\int_{\Sigma_\tau} \left( \partial_\tau \gamma_{ij}\pi^{ij}
+\partial_\tau\phi\,\pi-N{\cal H}-N^i{\cal H}_i\right)\sqrt{h}\right],
\end{equation}
where $\cal H$ and ${\cal H}_i$ are the Hamiltonian and momentum 
constraints respectively. The field equations imply that 
${\cal H}={\cal H}_i=0$, and furthermore the symmetry implies 
$\partial_\tau\phi=\partial_\tau \gamma_{ij}=0$. 
Therefore only the boundary term in (\ref{Ifull}) survives.

To evaluate the boundary term, we use the fact that the metric is 
static and asymptotically AdS, therefore at large $r$ we have
\begin{equation}
ds^2=f(r)d\tau^2+f(r)^{-1}dr^2+r^2d\Omega^2
\label{thesamemetric}
\end{equation}
where $f = 1-2GM/r-\Lambda r^2/3$.
For this metric $n_bu^a\nabla_a u^b= f^{-1/2}f'/2$, 
and subtracting the Euclidean AdS action from (\ref{Ifull}) 
we arrive at the following expression
\begin{equation}
I=\lim_{r\to\infty}\left({\beta r^2f'\over 4G}
-\frac{\beta_0 r^2 f_0'}{4G}\right),
\label{Ilim}
\end{equation}
where $\beta_0$ and $f'_0$ are the time-period and 
metric function of AdS space, and using \eqref{betaadsren}
the Euclidean action (\ref{Ilim}) becomes,
\begin{equation}
I=\beta M.
\end{equation}
If there is a horizon at $r=r_h$, then by properly treating the 
conical singularity as in \cite{GMW}, we get an additional area 
term contribution to the action
\begin{equation}
I=\beta M-\frac{1}{4G}{\cal A}_h.
\label{stbubble-action}
\end{equation}
This result generalises a previous result of Hawking and Horowitz, 
who found the same formula for the Euclidean action of static 
Einstein-matter solutions with $\Lambda=0$ 
and no conical singularities \cite{Hawking:1995fd}.
The $\Lambda\to0$ case can also be obtained using the
Gibbons-Hawking subtraction procedure described in 
\textsection \ref{actioncomputation}. The expression 
(\ref{stbubble-action}) is of course the same as
(\ref{schback}, \ref{adsschback}), but we have not assumed anywhere
for the static case that the bubble is in the thin-wall limit.

Solutions with a moving bubble wall $\left ( \tau(\lambda), R(\lambda)\right)$
break the time-translation symmetry of the full space-time, but the canonical 
method can still be used if the wall is thin and the geometries on both sides of the 
bubble wall, ${\cal M}_\pm$, individually possess the Killing vector, $\partial_\tau$.
Along with the contributions from ${\cal M}_\pm$ and the wall, $\cal{W}$, 
we will have an additional contribution from the conical singularity 
which can be dealt with by the methods of \cite{GMW}. 
The action splits into contributions from each of these parts
\begin{equation}
I_{\mathcal{M}_r}=I_-+I_++I_{\cal W} - \frac{{\cal A}_-}{4G},
\end{equation}
where ${\cal A}_-$ is the area of the black hole horizon
in the interior of the bubble.
In the thin wall limit $I_{\cal W} = \int_{{\cal W}}\sigma$, and
\begin{equation}
I_\pm=-\frac{1}{16\pi G}\int_{{\cal M}_\pm}{\cal R}\sqrt{g}
-\int_{{\cal M}_\pm}{\cal L}_m(g,\phi)\sqrt{g}
+\frac{1}{8\pi G}\int_{{\cal W}} K_\pm\sqrt{h},
\end{equation}
Hence, performing the same decomposition as in the static case 
we can cancel the bulk contributions and are left only with the boundary terms
\begin{equation}
I_\pm= -\frac{1}{8\pi G}\int_{\cal W}K_\pm \sqrt{h}
+ \frac{1}{8\pi G}\int_{\partial{\cal M}_\pm}n_{\pm b} u^a\nabla_a u^b\sqrt{h},
\end{equation}
where $\partial{\cal M}_+$ now includes the wall and the large 
distance boundary $\partial{\cal M}_r$. Using Israel's junction condition 
to relate the extrinsic curvatures on each side of the wall to the tension,
and inserting the normal to the wall (\ref{normal}), 
$n_bu^a\nabla_a u^b= \dot\tau f^{-1/2}f'/2$,
we reach our final result
\begin{equation}
I=\beta M_+-\frac{1}{4G}{\cal A}_- -\frac12\int_{\cal W}\sigma\sqrt{h}
-\frac{1}{16\pi G}\int_{\cal W}
\left(f_+'\dot{\tau}_+- f_-' \dot{\tau}_-\right)\sqrt{h},
\label{generalaction}
\end{equation}
The expression for the bounce action $B$ which governs 
the decay rate is obtained from $I$ by subtracting the 
background action without the bubble, $I_0$:
\begin{equation}
I_0=\beta M_+-\frac{1}{4G}{\cal A}_+.
\end{equation}
Therefore the tunnelling rate is determined by
\begin{equation}
B={{\cal A}_+\over 4G}-{{\cal A}_-\over 4G}-\frac12\int_{\cal W}\sigma\sqrt{h}
-\frac{1}{16\pi G}\int_{\cal W}
\left(f_+'\dot{\tau}_+- f_-' \dot{\tau}_-\right)\sqrt{h}\,,
\end{equation}
which is identical to \eqref{bounceaction}, after using the relation 
$f_+ {\dot\tau}_+ - f_- {\dot\tau}_- = -2{\bar\sigma}R$ 
for the wall integral.

\section{Higher dimensional instantons}
\label{higherD}

Here we briefly outline the simple higher dimensional instanton
model. Note this is only a crude calculation meant to give an 
indication of the effect of large extra dimensions, and is
problematic as an approximation to an actual Higgs tunneling
event due to the lack of an exact braneworld black hole solution,
as well as the fact that the Higgs should be confined to the brane,
hence the instanton should instead involve only a shift
in the cosmological constant on the brane. 
We take a higher dimensional black hole 
solution with different masses and cosmological constant
on each side of the wall, and compute how the branching ratio
changes with dimension. 

The equations of motion have the same schematic
form as \eqref{junctions}
\be
{\dot R}^2 = {\bar\sigma}^2 R^2 - {\bar f} 
+ \frac{(\Delta f)^2}{16{\bar\sigma}^2R^2}
\ee
but with ${\bar\sigma} = 4\pi G\sigma/(D-2)$ 
\cite{Gregory:2001xu,Gregory:2001dn},
and $f$ now 
the higher dimensional Schwarzschild potential:
\be
f = 1 - \frac{2\Lambda r^2}{(D-1)(D-2)} 
- \frac{16\pi G_DM}{(D-2) A_{D-2}r^{D-3}}
\ee
where $G_D$ is the higher dimensional Newton constant and
$A_{D-2} = 2 \pi^{\frac{D+1}{2}}/\Gamma[\frac{D+1}{2}]$.

Defining $\ell$, $\gamma$ and $\alpha$ in a similar fashion:
\be
\ell^2 = \frac{(D-1)(D-2)}{2\Delta\Lambda}\quad,\quad
\gamma = \frac{4{\bar\sigma}\ell^2}{1+4{\bar\sigma}^2\ell^2}
\quad , \quad \alpha^2 = 1 + \frac{2\Lambda_-\gamma^2}{(D-1)(D-2)}
\ee
and
\be
\beal
k_1 &= \frac{16\pi G_D}{(D-2)A_{D-2}} 
\left ( \frac{\alpha}{\gamma} \right ) ^{D-3} \left [
M_- + (1-\alpha)\frac{\Delta M}{2{\bar\sigma}\gamma}\right]\\
k_2 &= \frac{16\pi G_D}{(D-2)A_{D-2}} 
\left ( \frac{\alpha}{\gamma} \right ) ^{D-2} 
\frac{\Delta M}{4{\bar\sigma}}
\eeal
\ee
Then setting ${\tilde R} = \alpha R/\gamma$, 
${\tilde \lambda} = \alpha \lambda/\gamma$ gives the equation
of motion
\be
\left ( \frac{d{\tilde R}}{d{\tilde\lambda}}\right)^2 =
1 - {\tilde R}^2 - \frac{k_1+2k_2}{{\tilde R}^{D-3}}
- \frac{k_2^2}{{\tilde R}^{2(D-2)}}
\ee
which is of the same form as \eqref{rdot} albeit
with different exponents of $\tilde R$. We can use
the same procedure as before to find the static
solution and the dynamical bubble which removes the
black hole. The static action (which is what is needed
for the branching ratio) is, as before, the difference
in horizon areas:
\be
B_D = \frac{A_{D-2}}{4G} \left (
r_+^{D-2} - r_-^{D-2} \right)
\label{BDdef}
\ee
where $r_\pm$ are determined numerically, and the corresponding
tunneling rate is
\be
\Gamma_D = \sqrt{\frac{B_D}{2\pi}}\,
\frac{e^{-B_D}}{r_+}
\ee

Meanwhile the Hawking temperature of the higher dimensional
black hole is \cite{Myers:1986un,Kanti:2004nr}
\be
T_H = \frac{(D-3)}{4\pi r_+}
\ee
To estimate the decay rate of the black hole, we will assume that 
the main channel is due to emission of particles on the
brane \cite{Emparan:2000rs,Harris:2003eg}, leading to
\be
\Gamma_H \sim 3.6 \times 10^{-4} 
\frac{64\pi G_D}{A_{D-2} r_+^{D-1}}
\frac{(D-3)^4}{(D-2)}
\ee
hence a branching ratio of
\be
\frac{\Gamma_D}{\Gamma_H} \approx
550 \frac{M_+ r_+}{(D-3)^4} \sqrt{B_*} e^{-B_*}
\ee

\providecommand{\href}[2]{#2}
\begingroup\raggedright\endgroup

\end{document}